\documentclass[12pt]{article}
       
\usepackage{latexsym}
\usepackage{amssymb}
\usepackage{amsmath}

\def\bbbone {{\mathchoice {\rm 1\mskip-4mu l} {\rm 1\mskip-4mu l}
{\rm 1\mskip-4.5mu l} {\rm 1\mskip-5mu l}}}
\def\io {\overline{\imath}}
\def\jo {\overline{\jmath}}
\def\zo {\overline{z}}
\def\oe {\overline{1}}
\def\oz {\overline{2}}
\def\od {\overline{3}}
\def\ov {\overline{4}}
\def\psio {\overline{\psi}}

\def\pt {\partial_{\tau }}
\def\tr {\mbox{tr}}
\def \Ih {\hat{I}}
\def \ih {\hat{i}}
\def\alphat {\widetilde{b}}
\def\mt {\widetilde{m}}

\def\Acal {{\cal R}}
\def\Bcal {{\cal T}}

\begin{document}
\begin{titlepage}                   

\title{\Large \bf Boundary fermions and the plane wave}  

\author{{\Large Tako Mattik}\thanks{mattik@phys.ethz.ch}\\[4ex]   Institut f\"ur Theoretische Physik
\\ ETH Z\"urich\\CH-8093 Z\"urich\\ Switzerland}
\date{04.05.06}
\maketitle \abstract{We construct branes in the plane wave background under   the inclusion of fermionic boundary fields.  The resulting deformed boundary conditions in the bosonic and fermionic sectors give rise to  new integrable and supersymmetric branes  of type $(n,n)$.  The extremal  case  of the  spacetime filling $(4,4)$-brane is  shown to be maximally spacetime  supersymmetric. }
\thispagestyle{empty}
\end{titlepage}

\noindent                     
\tableofcontents          
\section{Introduction}
In recent  years, fermionic boundary fields have been used  in  a number of different field theoretical settings. Soon after the introduction of boundary fermions for the massive Ising model defined on a manifold with boundaries in \cite{ghoshal} in the context of integrable boundary models, Warner initiated   in \cite{warner1} their application in the study  of ${\cal N}=(2,2)$ supersymmetric   Landau-Ginzburg models.   
In this case,  terms corresponding to   the boundary fermions are chosen to cancel a  generically nonzero boundary contribution  (the  so called   Warner term) in the supersymmetry variation of the  bulk Lagrangian in the presence of D-branes. In the context of integrable boundary field theories similar  ideas have  thereafter been used in \cite{nepo1,  nepo2} to construct boundary  conditions compatible with supersymmetry and integrability for extensions of the sine-Gordon model.   In the string community  they  finally led to the study of matrix factorisations and the relation of  D-brane physics and coherent sheaves, compare  for example  with \cite{govind, kapustin2, brunner1} as initial references. \\[1ex]  Although we will use a   Lagrangian analogously   to those appearing in    \cite{nepo2, brunner1, kapustin2},  we want to mention that  similar  boundary fermionic fields also appeared in constructions aiming at   nonabelian extensions of  
Dirac-Born-Infeld D-brane descriptions.  Going back to \cite{marcus},  the boundary fermions are in this case  interpreted as representing Chan-Paton factors.    A discussion along these lines  in the context of pure spinors  is presented in  \cite{berkovits2} and   a treatment using the Green-Schwarz formulation is to be found \cite{howe1}.\\[2ex]
The relation of  plane wave physics to  boundary fermionic fields   in the context of   ${\cal N}=2$ supersymmetry  is most easily derived  from  the  work  of Maldacena and Maoz in  \cite{maldacena} on   nontrivial Ramond-Ramond  type II B supergravity backgrounds, chosen to  preserve at least 4 spacetime supersymmetries. For a flat transverse space these backgrounds  of pp-wave structure are  exact superstring solutions \cite{berkovits}  and in this case  parametrised by a single  holomorphic function. In the corresponding worldsheet theory, given by a ${\cal N}=(2,2)$ supersymmetric Landau-Ginzburg model, this function becomes the worldsheet superpotential $W({\bf z})$.  For further  applications  of   methods from  \cite{berkovits} to comparable backgrounds as constructed in \cite{maldacena},  see  for example \cite{bonelli2}.\\ The choice of a trigonometric superpotential  $W(z)\sim \cos z$  in the solutions  of \cite{maldacena} leads to the integrable ${\cal N} =2$  supersymmetric sine-Gordon model on the worldsheet, whereas the exponential $W(z)\sim e^z$ gives rise to the ${\cal N}=2$ Liouville theory. In the context of boundary fermions these theories have been discussed  in \cite{nepo2, baseilhac, mattik2} and \cite{ahn}.  Using an approach in this spirit, but see also the Lagrangians defined in \cite{brunner1, kapustin2},  we will be interested  in this paper in the situation described by  the superpotential \begin{eqnarray}\label{start1} W({\bf z})=-im\sum_{j=1}^4 (z^j)^2 .\end{eqnarray}  As pointed out   for example in \cite{maldacena},  one reobtains from (\ref{start1})  the situation of strings in the maximally supersymmetric type II B  plane wave background from \cite{blau},  described by the metric  \begin{equation} \label{start2} ds^2=2 dX^+ dX^--m^2 X^I X^I dX^+dX^+ + dX^IdX^I \end{equation} and the nontrivial five-form components \begin{equation} F_{+1234}=F_{+5678}=2m.\end{equation}  The closed string theory in this  background  was first solved in \cite{metsaev1} and attracted a substantial amount of interest in particular after the appearance of \cite{berenstein} linking the  string theory on (\ref{start2}) to the general study of the AdS/CFT correspondence from \cite{malda2}. For reviews of this field  see \cite{reviewpp}.\\[2ex]
Branes in the plane wave background  of \cite{blau} have been studied in a number of papers from different point of views. We will briefly review them and  the classification of the maximally supersymmetric branes into class I and II branes from \cite{skentay2, gabgre2}  in section \ref{braneintro}  where we will also point to the corresponding literature. \\ Here we only  want to mention that all the maximally supersymmetric branes known for this  background are also integrable, that is,  they preserve the integrable structure of the closed string theory in the sense of \cite{ghoshal}.  Handling   a free theory, relatively little attention is usually payed to this point. However, the inclusion of boundary fermions  modifies  this situation as they generically  give  rise to an interacting boundary field theory which is in most cases  also  incompatible with integrability.  The requirement  of conserved higher spin currents  in the boundary theory will lead to strong constraints on  admissible boundary couplings.   It is worth mentioning that the massive Ising model,  appearing  as the  fermionic part of the plane wave worldsheet theory  of (\ref{start2}), has been intensively discussed in the literature on integrable (boundary) models, see for example  \cite{ghoshal, chatterjee} and references therein. \\[1ex] 
 From the point of view of the ${\cal N}=2$ worldsheet  supersymmetry,  branes  in Maldacena-Maoz backgrounds  have been studied in \cite{hikida},  building on  the work of \cite{vafa2}. In this case  the ``Warner problem" is avoided  by the choice of   particular (oblique) orientations of the Neumann directions,  implying a vanishing  Warner term. The oblique branes of \cite{hikida}  for the particular  plane wave  background have been studied along the lines of \cite{start, skentay2, gabgre2} in \cite{oblique}. \\ 
  In this paper we discuss   branes   beyond the restrictive setting of \cite{hikida} by  aiming first of all  at  integrable branes  with   preserved   ${\cal N}=2$ (worldsheet)  supersymmetry under the inclusion of fermionic boundary fields.  In the classification of \cite{skentay2, gabgre2} the  new branes are  all of type  $(n,n)$ and  as a main result, the limiting case of the  spacetime filling $(4,4)$-brane with only Neumann directions in the transverse space  is found to be maximally spacetime  supersymmetric.  This is  in analogy to the other limiting case of the $(0,0)$ instanton from \cite{skentay2, gabgre2}, with which it also shares   analogous boundary state overlaps.\\    The  bosonic boundary conditions  of the new branes are  expressible as a standard  coupling to a nonzero longitudinal  flux ${\cal F}_{+I}$. The fermionic  bulk and boundary fields, on the other hand, are first of all  determined due to a coupled system of differential equations on the boundary.  The on-shell elimination of the  boundary fermions from this system leads to  an expression for the bulk field boundary conditions in terms of a   linear differential equation in the boundary parametrising coordinate $\tau$.  As an interesting result, the boundary fermions can finally be expressed as a function  of the bulk fermionic fields  without  including additional degrees of freedom. In the quantum theory  the  corresponding  expressions also  correctly reproduce  the required   quantum mechanical anticommutation relations for the boundary fields. \\[2ex] This paper is organised as follows. In the starting section \ref{landau} we collect  background information on  the plane wave theory formulated as a Landau-Ginzburg model and state  the relation between the ${\cal N}=(2,2)$ worldsheet supercharges and the maximal spacetime supersymmetry from \cite{blau, metsaev1}.  After briefly reviewing  branes in the plane wave theory  in section \ref{braneintro},   we start in section \ref{boundaryintro} our  study of boundary fermions in the context of plane wave physics and derive the conditions for integrable and ${\cal N}=2$ supersymmetric branes.  The branes  solving  these conditions are then studied in detail in section \ref{openstring}  by constructing and quantising the corresponding open string theory.  In the subsequent section \ref{spacesusy} we conduct a discussion    using   boundary states, leading in particular to  a study  of  preserved spacetime supersymmetries in the presence of boundary fermions. Here we also briefly suggest,  following \cite{skentay1},  how to realise the  deformed Neumann  boundary conditions  in the bosonic sector by nonzero longitudinal  fluxes. In the final section \ref{openclose},  the equivalence of the open and closed string constructions is discussed along the lines of \cite{gabgre, gabgre2} by establishing the equality of certain open string partition functions with the corresponding closed string boundary state overlaps.  Certain technical details are collected in the appendices.

  \section{The plane wave as a Landau-Ginzburg model}
\label{landau} In this section we collect some information  about the  worldsheet theory for strings in the maximally  supersymmetric plane wave background of  \cite{blau}  formulated as a ${\cal N}=(2,2)$ supersymmetric  Landau-Ginzburg model. In particular,  we mention  the relation between the Landau-Ginzburg   and   Green-Schwarz fermions along the lines of \cite{maldacena}. This will especially  also lead to  expressions for the ${\cal N}=(2,2)$ supercharges as linear combinations of the spacetime supersymmetries from \cite{blau, metsaev1}.  As these results are crucial for the later sections,  we supply some additional details in the appendix \ref{eta}. Our conventions for  Landau-Ginzburg models are those   summarised for example  in \cite{ hikida}.\\[2ex]
From  a Landau-Ginzburg model with the general component Lagrangian 
\begin{eqnarray} \nonumber  {\cal L}_{\mbox{bulk}}  &=& \frac{1}{2} g_{j\jo} \left(\partial_+ z^j \partial_- \zo^{\jo}+\partial_+ \zo^{\jo}\partial_- z^j+i \overline{\psi}^{\jo}_+ \stackrel{\leftrightarrow}{\partial}_- \psi^j_++i\overline{\psi}_-^{\jo}\stackrel{\leftrightarrow}{\partial}_+ \psi^j_-  \right)\\ &\;&\; -\frac{1}{2}\partial_i\partial_j W({\bf z})\;  \psi^i_+\psi^j_- - \frac{1}{2}\partial_{\io}\partial_{\jo}\overline{W}(\overline{\bf z})\; \overline{\psi}^{\io}_-\overline{\psi}_+^{\jo}-\frac{1}{4}g^{i\jo} \partial_iW({\bf z})\; \partial_{\jo}\overline{W}(\overline{\bf z}), \label{lagragen}    \end{eqnarray}   the  plane wave theory from \cite{blau, metsaev1}  is obtained,  following \cite{maldacena}, by  setting the superpotential  as mentioned in the introduction to 
\begin{equation}\label{superpot1} W({\bf z})=-im\sum_{i=1}^4 (z^i)^2;\;\;\;\;\;\;  \overline{W}(\overline{{\bf z}})=im\sum_{\jo=1}^4 (\zo^{\jo})^2.\end{equation} This choice gives rise to the equations of motion  
  \begin{equation} \label{eomboslg} \left(\partial_+\partial_-+m^2\right)z^i=0= \left(\partial_+\partial_-+m^2\right)\zo^{\io}\end{equation} for the bosons and \begin{eqnarray} \nonumber 0=  \partial_- \psi^{j}_++m\overline{\psi}^{\jo}_- \;\;\;&\;&\;\;\; 0= \partial_- \overline{\psi}^{\jo}_++m{\psi}^{j}_-   \\  0= \partial_+ \psi^{j}_--m   \overline{\psi}^{\jo}_+\;\;\; &\;&\;\;\; 0= \partial_+ \overline{\psi}^{\jo}_--m {\psi}^{j}_+.   \label{eomferlg} \end{eqnarray} for the fermions.\\[2ex]  The relation between the fermions in (\ref{lagragen}) and the standard Green-Schwarz fields $S, \widetilde{S}$ was pointed out in \cite{maldacena}  and is given by  \begin{eqnarray}\label{identi1} S^a &=& \psi_-^i\Gamma^{ab}_i \eta^b +\overline{\psi}^{\io}_- \Gamma^{ab}_{\io}\eta^{*b}\\ \label{ident2} \widetilde{S}^a&=&\psi_+^i\Gamma^{ab}_i \eta^b +\overline{\psi}^{\io}_+ \Gamma^{ab}_{\io}\eta^{*b}\end{eqnarray}   with a constant spinor $\eta$ fulfilling  \begin{eqnarray}\label{eta1} 0=\Gamma_{\io}\eta;\;\;\; \eta^*\eta=1;\;\; \Pi \eta=-\eta^*. \end{eqnarray}
The (new)  Majorana type requirement $\Pi \eta=-\eta^*$   contains  the real matrix  
  \begin{eqnarray} \Pi = \gamma^1\gamma^2\gamma^3\gamma^4 \label{Pi}\end{eqnarray} from \cite{metsaev1} and is   consistent  due to $\Pi^2=1$.   It is chosen to determine  an up to a sign unique spinor  $\eta$ and it  correctly reproduces  the equations of motion 
   \begin{equation}\label{eomgs} \partial_+ S=m\Pi \widetilde{S};\;\;\;\; \partial_- \widetilde{S}=-m\Pi S\end{equation}  for the GS fields by  starting from (\ref{eomferlg}).
  \\ We will further discuss these identifications in the appendix \ref{eta} and close this section   by stating the relation between the ${\cal N}=(2,2)$ worldsheet supercharges and the spacetime supersymmetries as derived in \cite{metsaev1}. Using  the spacetime charges  in the  conventions of \cite{gabgre2}, the required  identifications  are given by \begin{eqnarray}\label{q1} \frac{Q_+}{\sqrt{2p^+}} = \eta^* \widetilde{Q} = -\eta \Pi \widetilde{Q} \;\;\; &\;& \;\;\;  \frac{Q_-}{\sqrt{2p^+}} = \eta^* Q = -\eta \Pi Q \\ \label{q2}
\frac{\overline{Q}_+}{\sqrt{2p^+}} = \eta \widetilde{Q} = -\eta^* \Pi \widetilde{Q} \;\;\; &\;& \;\;\;  \frac{\overline{Q}_-}{\sqrt{2p^+}} = \eta {Q} = -\eta^* \Pi {Q} \end{eqnarray} and we again defer a derivation to the appendix. 
  
\section{Branes in the plane wave background}
\label{braneintro}
In this section we briefly review  the classification of (maximally) supersymmetric branes in the plane wave background from \cite{skentay2, gabgre2} to explain  the context  of  our subsequent constructions. As mentioned in the introduction, soon after the solution of  the closed string theory in the plane wave background  from \cite{blau} in \cite{metsaev1},  branes in this background have been studied in  a significant number of papers. These  discussions include various approaches, for example the use of open strings, closed string boundary states or  geometric methods like probe brane settings. Starting with the papers \cite{start},  details  about branes in the type II B plane wave background were derived in  \cite{skentay1, skentay2, skentay3, gabgre,  gabgre2, oblique, chu, cha1, mattik, lee3,  pwbranes} and  related settings are   discussed  for example in \cite{related}. \\[1ex]  Following in particular the flat space treatment  in  \cite{gregut}, the (maximally) supersymmetric branes in the plane wave background have been classified in \cite{skentay2, gabgre2} by using the spinor matrix \begin{eqnarray} M=\prod_{I\in{\cal N}} \gamma^I.\end{eqnarray}  The product  is  understood to span over the Neumann directions and the matrix $M$ appears in the standard fermionic boundary conditions.  \\ Branes of class I are characterised by  \begin{equation} M\Pi M\Pi =-1\end{equation}  and the maximally supersymmetric branes of this type are of  structure  $(r,r+2)$, $(r+2, r)$ with $r=0,1,2$.  Here the notation $(r,s)$ from \cite{skentay1} labels  the brane's orientation with respect to the $SO(4)\times SO(4)$-background symmetry. \\ For class II   branes  one has  \begin{equation} M\Pi M\Pi =1\end{equation}  and the known  maximally supersymmetric branes in this class are    the $(0,0)$ instanton and the $(4,0), (0,4)$ branes from \cite{skentay2, gabgre2}. \\ One of the  main results  in  this paper  is the construction of a maximally supersymmetric class II brane of type $(4,4)$   with deformed fermionic boundary conditions, originating from   the inclusion of  boundary  fermions.  All our new  branes  will be   class II  branes of type $(n,n)$. For $n=1,\dots 4$ one can find alternative constructions without boundary excitations  in \cite{cha1}.  In this case,  however,  there are only 4 conserved supersymmetries throughout. We defer a discussion of this setting to section \ref{limitcompare}.\\  It is worth pointing out  that the inclusion  of a boundary magnetic field as discussed in \cite{mattik, lee3} allows to construct maximally supersymmetric branes which interpolate between the  class I  $(2,0)$ / $(4,2)$-branes and the class II $(0,0)$-instanton and the  $(4,0)$ brane, linking the two families in a natural way. \\[2ex]  
As mentioned in the introduction, we will begin by   focussing   on  ${\cal N}=2$ supersymmetric settings   in conjunction with a preserved integrable structure following \cite{warner1}.   Integrability is a shared feature of all the maximally  supersymmetric branes in the plane wave background\footnote{This can be proven by applying  the methods from \cite{ghoshal} to be briefly   mentioned in the appendix \ref{appintegrability}. For the $(0,0)$-instanton  as a particular $(n,n)$-brane   this result will be established in due course.}, but is generically lost in the presence of boundary fermions with general couplings to the bulk fields. The enforcement of integrability to be discussed in the next section  will determine these boundary couplings up to  constant parameters. \\[1ex] 
As we will conduct  a study of the  previously mentioned settings from an  open and    closed string perspective, it is important to notice  that  the  standard light-cone gauge condition gives rise to branes of different nature in these two  sectors. In the open sector the light-cone directions  in the standard gauge are of Neumann type, whereas they become Dirichlet like   in the closed string sector, leading therefore to instantonic boundary states, compare with \cite{gregut}.  As explained in \cite{gabgre, gabgre2, skentay3}, one has to apply different light-cone gauge choices in the two sectors to allow for a direct comparison. Due to this,  gauge dependent quantities like the mass $m$  appearing in (\ref{superpot1}) take on different values in the two cases. We will discuss the relation between the closed string  constants $m, b^i, k^i$ and their open string correspondents $\mt, \alphat^i, \widetilde{k}^i$   along the lines of \cite{gabgre, gabgre2} in section \ref{transverseflux}. \\ In the  following  sections  two directions combined to a complex variable  \begin{eqnarray} z^j= x^j+ix^{j+4}\;\;\;\; j=1,\dots 4\end{eqnarray} are always chosen to have the same type of boundary conditions.  For later convenience we furthermore define sets ${\cal D}_-$, ${\cal N}_-$ containing the Dirichlet and Neumann directions ranging in  $r=1,\dots, 4$  and correspondingly      ${\cal D}_+$, ${\cal N}_+$ with elements in  $r=5,\dots, 8$.

\section{Boundary  fermions: Supersymmetry and integrability}\label{boundaryintro}
In this section we will start to  construct  branes  in the plane wave background under the inclusion of boundary fermionic fields. In a first step,  we define a suitable  boundary Lagrangian and derive the corresponding boundary conditions   for the bulk fields and the equations of motion for the boundary fermions.\\  Using these conditions, we can thereafter calculate the determining equations for the  boundary fields under the requirement of conserved ${\cal N}=2$ supersymmetry and integrability in the boundary theory. Further information about the integrable structure and calculational details omitted in this section  can be found in the appendix \ref{appintegrability}. 

\subsection{Boundary conditions}
 By mildly extending  the  boundary Lagrangians defined in \cite{ghoshal,warner1, nepo2},  see also \cite{brunner1, kapustin2},  to include matrix valued boundary fields,  we will work subsequently with the real Lagrangian 
 \begin{eqnarray} \nonumber {\cal L}^{\sigma=\pi}_{\mbox{boundary}}  &=&\frac{i}{2}g_{j\jo}\left(e^{-i\beta} \psi^j_-\psio^{\jo}_+   -e^{i\beta} \psi^j_+\psio^{\jo}_- \right)  -\frac{i}{2}\; \mbox{tr}\left[A\stackrel{\leftrightarrow}{\partial}_\tau A^\dagger\right]+B({\bf z,\zo})  \\  \nonumber &\;\;& +\frac{i}{2} \;\tr\left[\partial_{\jo} F^\dagger(\overline{\bf z}) A^\dagger+\partial_{\jo} G^\dagger(\overline{\bf z}) A\right]\left(\psio_+^{\jo}+e^{i\beta}\psio_-^{\jo}\right)\\ \label{boundlagra} &\;\;&+\frac{i}{2}\; \tr\left[\rule{0mm}{2mm}
\partial_jG({\bf z})A^\dagger+\partial_j F({\bf z})A\right]\left(\psi_+^j+e^{-i\beta}\psi_-^j\right)\end{eqnarray}  defined along the Neumann directions at the boundary $\sigma=\pi$. The square matrix $A=(a_{rs})$ contains the boundary fermions and $F,G$ are matrix valued functions of the bosonic bulk fields evaluated on  the boundary.\\
The boundary conditions along the Neumann directions deducing  from the variation of (\ref{lagragen}) and    (\ref{boundlagra})   are found to be \begin{eqnarray} \label{bcond1} \partial_\sigma z^j &=& g^{j\jo}\left( \partial_{\jo} B + i\; \tr\left[\partial_{\io}\partial_{\jo} F^\dagger A^\dagger+\partial_{\io}\partial_{\jo} G^\dagger A\right]\overline{\theta}_+^{\io} \right)
\\   \label{bcond2} \partial_\sigma \zo^{\io} &=& g^{j\jo}\left(\partial_j B + i\; \tr\left[\rule{0mm}{2mm}
\partial_i\partial_j GA^\dagger+\partial_i\partial_j FA\right]\theta^i_+\right)\\ \label{bcond3}
\theta^j_- &=& \frac{1}{2}g^{j\jo}\; \tr \left[\partial_{\jo}F^\dagger A^\dagger+\partial_{\jo}G^\dagger A\right]\\   \label{bcond4}\overline{\theta}^{\jo}_- &=& \frac{1}{2} g^{j\jo} \; \tr\left[ \partial_jG A^\dagger+\partial_j F A\right]
\\  \label{bcond5}  \pt A &=& \partial_{\jo}F^\dagger \overline{\theta}^{\jo}_++\partial_jG \theta^j_+ \\  \label{bcond6} \pt A^\dagger &=& \partial_{\jo}G^\dagger \overline{\theta}^{\jo}_++\partial_jF \theta_+^j\end{eqnarray} 
which is understood to be evaluated at $\sigma =\pi$ throughout.  We have furthermore used the convenient combinations 
 \begin{eqnarray} \nonumber \theta_+^i= \frac{1}{2} \left(\psi^i_++e^{-i\beta}\psi^i_-\right)\;\;\; &\;&\;\;\; \overline{\theta}^{\io}_+=\frac{1}{2} \left(\overline{\psi}^{\io}_++e^{i\beta}\overline{\psi}^{\io}_-\right)\\  \theta^i_-= \frac{1}{2} \left(\psi^i_+-e^{-i\beta}\psi^i_-\right)\;\;\; &\;&\;\;\; \overline{\theta}^{\io}_-=\frac{1}{2} \left(\overline{\psi}^{\io}_+-e^{i\beta}\overline{\psi}^{\io}_-\right)  \end{eqnarray} for the bulk fermions.  By setting \begin{eqnarray} {\cal L}^{\sigma=0}_{\mbox{boundary}} = -{\cal L}^{\sigma=\pi}_{\mbox{boundary}}\end{eqnarray} one obtains functionally the same boundary conditions at $\sigma=0$ as derived beforehand for $\sigma=\pi$ with, however, possibly  different matrices  $F, G$ at the two boundaries.  Although the constraints on $F$ and $G$ to be derived below are also valid in the case of different boundary fields, we will focus on the case of equal boundary conditions up to  different choices for $\beta$,  corresponding to brane / antibrane configurations.\\[1ex]  Along the Dirichlet directions   we will use the standard  boundary conditions  as for example discussed in \cite{vafa2}. These are in particular independent of the previously introduced boundary fermions and  read  explicitly \begin{eqnarray} \label{bcond7} z^i=y^i_{0,\sigma} ;\;\; \overline{z}^{\io} =\overline{y}^{\io}_{0,\sigma}\\ \label{bcond8}  0= \theta_+^i;\;\; 0=\overline{\theta}^{\io}_+.\end{eqnarray}  All fields are again understood to be evaluated at $\sigma=0,\pi$. 
 
 \subsection{B - type supersymmetry} \label{branesusy2}
 As explained in section \ref{braneintro},  we  consider first of all  boundary conditions  by aiming at branes with two conserved B - type supersymmetries.  As pointed out  in \cite{ghoshal} in a  different context,  the open string conservation of quantities deducing from local   conserved fluxes  amounts to the time independency  of (in our case) the following combinations   \begin{eqnarray}\label{qboundary1} Q& =& \overline{Q}_+ + e^{i\beta}\overline{Q}_-+\Sigma_\pi(\tau)-\Sigma_0(\tau)\\ \label{qboundary2}  Q^\dagger &=& Q_++e^{-i\beta}Q_-+\overline{\Sigma}_\pi(\tau)-\overline{\Sigma}_0(\tau)\end{eqnarray} with  generically nonzero ({\em local}) contributions of    boundary fields $\Sigma_{\sigma}(t)$ at $\sigma=\pi$ and $\sigma=0$.\\  By using the supercurrents (\ref{currents1}) and (\ref{currents2}) presented in the appendix  \ref{eta},  the quantities  (\ref{qboundary1}) and (\ref{qboundary2})  are time independent in case of  \begin{eqnarray}\label{fluxw3}  0&=&\left. \overline{G}_+^1+e^{i\beta} \overline{G}^1_-\right|_{\sigma=\pi} -\dot \Sigma_\pi(\tau)\\  \label{fluxw4}  0&=&\left. \overline{G}_+^1+e^{i\beta} \overline{G}^1_-\right|_{\sigma=0} -\dot \Sigma_0(\tau). \end{eqnarray}
  Along the Dirichlet directions these conditions are trivially fulfilled with the boundary conditions (\ref{bcond7}) and (\ref{bcond8}) together with a vanishing  field $\Sigma_\sigma$ along these directions.  In the   case of  Neumann directions  with  boundary conditions (\ref{bcond1})-(\ref{bcond6}) the situation is more interesting.   For a  single Neumann direction   the solution to (\ref{fluxw3}) and (\ref{fluxw4})   is discussed in detail in  \cite{mattik2} and that  treatment  extends immediately to the present situation including matrix valued boundary fields.  Suppressing the calculational details,  we   obtain the conditions 
  \begin{eqnarray}\label{susycond1} B&=& \frac{1}{2} \;\tr\left[GG^\dagger+FF^\dagger\right]+const \\  W &=& i e^{-i\beta} \;\tr\left[FG\right]+const  \label{susycond2}. \end{eqnarray}  The second equation (\ref{susycond2})  is   understood  to be valid along the Neumann directions only. 
 For the local boundary field $\Sigma_\pi$ we furthermore have \begin{eqnarray} \Sigma_\pi (\tau) = -2 g_{j\jo} \overline{\theta}^{\jo}_-z^j+\;\tr\left[\left(z^j\partial_jF-F\right)A+\left(z^j\partial_jG-G\right)A^\dagger  \right],  \label{susycond3} \end{eqnarray}  compare again with \cite{mattik2}.

 \subsection{Integrability}\label{braneintegrability}
Although arbitrary boundary fields obeying (\ref{susycond1}) and (\ref{susycond2}) already give rise to ${\cal N} =2$ supersymmetrical settings, we are here interested  in the more restricted case of \emph{integrable} boundary conditions, that is,  branes which also respect the integrable structure present in the bulk theory.    As explained in section \ref{braneintro},  all known maximally supersymmetric  branes in the plane wave theory  are actually also integrable.   By the inclusion   of  boundary fields  as  in (\ref{boundlagra}),  this integrability conservation  is a priori   no longer guaranteed and leads, if enforced,  to further constraints on admissible boundary conditions. \\[1ex]  In this section we will give the explicit expression of two higher spin bulk  currents  and state the conditions for their conservation in the presence of  boundaries. This  conservation gives strong evidence for the integrability of the boundary theory. To further underpin the  actual presence of  such a structure one might use   the explicit mode expansions to be derived in the next section and compare them with the requirements derived in \cite{ghoshal} for integrable boundary field theories. We will  briefly comment on this in the appendix \ref{appintegrability}.\\ Local conserved higher spin currents for the  massive Ising model were written down in \cite{zamo1}.   Here we  focus on   combinations  which, for a single Neumann direction, appear as limiting cases of the first nontrivial higher spin currents in the ${\cal N}=2$ sine-Gordon model.
 We defer a more detailed discussion  of this point to the   appendix  \ref{appintegrability} where we also supply the infinite series of conserved fluxes from \cite{zamo1}.\\  In manifestly real form the currents  of present interest are given by  \begin{eqnarray} \label{influx1}T_4 &=& g_{i\io}\left(\; \partial_+^2\zo^{\io}\;\partial_+^2 z^i+\frac{i}{2}\partial_+\psio_+^{\io}\;\partial_+^2\psi_+^i-\frac{i}{2}\partial^2_+\psio_+^{\io}\;\partial_+\psi_+^i\right)\\  \label{inftlux2}\theta_2 &=& g_{i\io}\left(-m^2 \partial_+\zo^{\io}\;\partial_+ z^i -\frac{im^2}{2} \psio_+^{\io}\;\partial_+\psi_+^i+\frac{im^2}{2} \partial_+\psio_+^{\io}\;\psi_+^i\right)\end{eqnarray}
and  \begin{eqnarray} \label{intflux3}\overline{T}_4 &=& g_{i\io}\left( \partial_-^2\zo^{\io}\;\partial_-^2 z^i+\frac{i}{2}\partial_-\psio_-^{\io}\;\partial_-^2\psi_-^i-\frac{i}{2}\partial^2_-\psio_-^{\io}\;\partial_-\psi_-^i\right)\\ \overline{\theta}_2 &=& g_{i\io}\left(-m^2 \partial_-\zo^{\io}\;\partial_- z^i -\frac{im^2}{2} \psio_-^{\io}\;\partial_-\psi_-^i+\frac{im^2}{2} \partial_-\psio_-^{\io}\;\psi_-^i\right) \label{intflux4}\end{eqnarray} and fulfil  on-shell
\begin{eqnarray}\label{intflux5} \partial_-T_4=\partial_+\theta_2;\;\;\;\partial_+\overline{T}_4=\partial_-\overline{\theta}_2.\end{eqnarray} In the bulk theory both fluxes give rise to conserved spin 3 operators.
The conservation of a suitable combination of the previous operators  in the presence of boundaries  is discussed in the appendix.  There the  conditions for  integrability are  found to be 
 \begin{eqnarray} \nonumber  \partial_i\partial_j\partial_k B=0 \;\; &\;&\;\; \partial_{\io}\partial_{\jo}\partial_{\overline{k}}B = 0\\    \label{condint1}\partial_i\partial_j\partial_{\overline{k}}B=0 \;\; &\;&\;\; \partial_{\io}\partial_{\jo}\partial_k B =0\end{eqnarray} for the boundary potential and  \begin{eqnarray} \nonumber 0 &=& \tr\left( \partial_i\partial_j G A^\dagger+\partial_i\partial_jF A\right)\\  \label{condint2}  0 &=& \tr\left( \partial_{\io}\partial_{\jo} G^\dagger  A+\partial_{\io}\partial_{\jo}F^\dagger  A^\dagger\right)\end{eqnarray} for the matrices $F$ and $G$.\\[3ex]  Having presented  the  conditions for ${\cal N}=2$ supersymmetry   (\ref{susycond1}), (\ref{susycond2})  in the last section and for integrability  in (\ref{condint1})  and (\ref{condint2}), it is now straightforward to write down the corresponding solutions.
 They are given by  \begin{eqnarray}\label{sol1} F=A_i z^i+C\;\;\;\;\; G=B_i z^i+D\end{eqnarray}  along the Neumann directions with \begin{eqnarray} \label{sol2} \tr\left(A_iB_j\right) = -e^{i\beta}\mt\delta_{ij}\;\;\; \tr\left(A_iD+B_iC\right)=0.\end{eqnarray}   The resulting  boundary potential becomes   up to an irrelevant constant \begin{eqnarray}\label{sol3} B ({\bf z, \overline{z}})=\frac{1}{2} \tr\left(A_iA^\dagger_{\jo} +B_iB^\dagger_{\jo}\right)z^i\zo^{\jo}+\tr \left(A_iC^\dagger +B_i D^\dagger \right) z^i+\tr \left(CA^\dagger_{\io}+DB^\dagger_{\io}\right)\zo^{\io} , \end{eqnarray}  again extending only along the Neumann directions. 
 
\section{The open string with boundary fermions} \label{openstring}
In this section we present  a detailed  discussion  of $(n,n)$-branes with $n=0,\dots, 4$ from an open string point of view  by enforcing  Neumann boundary conditions as introduced in the last section. Using the equations of motion for the boundary fermions we can eliminate these  extra fields from the remaining  boundary conditions.  Although the resulting  boundary conditions  on the fermionic bulk fields  differ clearly from  the standard settings, the corresponding  solutions can be found  and quantised by  standard methods.\\ As stated in the introduction, the boundary fermions can be expressed in terms of the bulk fields restricted to the boundary without including additional degrees of freedom. We explain in detail how this solutions reproduces the expected anticommutators of the boundary fermions in the quantum theory. The section closes with a derivation of the ${\cal N}=2$ superalgebra of the boundary theory. These results will be needed in the discussion of the open-closed duality in section \ref{openclose}.  \\[1ex]  For the boundary fields appearing in the Neumann directions  we will work  with   a particular solution of  type (\ref{sol1})  given by \begin{eqnarray}\label{sol4} F=\text{diag}(A^{\ih}z^{\ih}+C^{i});\;\; G=\text{diag}(B^{\ih}z^{\ih}+D^{i})\end{eqnarray} with no sum over  hatted indices. The solution (\ref{sol4})  allows us to treat the fields along any complex direction $z^i$ separately and construct $(n,n)$-type branes for  all $n$  in a single approach. \\ We will consider only strings spanning between branes with the same type of boundary fields and  restrict the parameter $\beta$ appearing in (\ref{boundlagra})  to the values $0$ and $\pi$ corresponding to  brane or  antibrane settings. The latter will again be  needed  in section \ref{openclose}.  The more general situation of $\beta\in (0,\pi)$ can be dealt with with the methods explained in \cite{mattik} in the context of boundary magnetic fields in  the plane wave background.\\[2ex]
For future reference we note here  the most general solutions to the equations of motion (\ref{eomboslg}) and (\ref{eomferlg}) which read in a real basis  \begin{eqnarray}\label{eom1} 0=\left(\partial_+\partial_-+\mt^2\right)X^s \end{eqnarray} for the bosons with $s=1,\dots 8$  and  \begin{eqnarray}\label{eom2} \partial_- \psi_+^t = -\mt\psi_-^t &\;\;& \partial_+\psi_-^t = \mt\psi_+^t \\  \label{eom3} \partial_-\psi_+^{t+4} =+\mt\psi_-^{t+4}&\;\;&\partial_+\psi_-^{t+4}=-\mt\psi_+^{t+4}\end{eqnarray} for the fermions  with $t=1,\dots 4$. \\  The fermionic fields along the $s=5,\dots 8$ directions  are   obtained from those along the  $s=1,\dots,4$ directions by interchanging $\mt\leftrightarrow -\mt$, reflecting the different eigenvalues of the matrix $\Pi$ introduced in section \ref{braneintro}. \\ Following \cite{gabgre2}, the most general  solutions  to (\ref{eom1})-(\ref{eom3}) are given by   \begin{eqnarray}\nonumber  X^s(\tau,\sigma)&=& C^s\sin(\mt\tau)+\widetilde{C}^s\cos(\mt\tau)+D^s\cosh(\mt\sigma)+\widetilde{D}^s\sinh(\mt\sigma)\\ &\;\;& +i\sum_{\stackrel{n,\omega_n\neq 0}{\omega_n^2=n^2+\mt^2}}\frac{1}{\omega_n}\left(a^s_n e^{-i(\omega_n\tau-n\sigma)}+\widetilde{a}^s_ne^{-i(\omega_n\tau+n\sigma)}\right) \label{allgemein1}\end{eqnarray} 
and \begin{eqnarray} \nonumber \psi^t_+(\tau,\sigma) &=& -\phi^t\sin(\mt\tau)+\widetilde{\phi}^t\cos(\mt\tau)+\widetilde{\psi}^t\cosh(\mt\sigma)+\psi^t\sinh(\mt\sigma)\\&\;\;& + \sum_{\stackrel{n,\omega_n\neq 0}{\omega_n^2=n^2+\mt^2}}c_n\left(\widetilde{\psi}^t_n e^{-i(\omega_n\tau+n\sigma)}-\frac{i}{\mt}(\omega_n-n)\psi_n^t e^{-i(\omega_n\tau-n\sigma)}\right)  \label{allgemein2}\\ \nonumber \psi^t_-(\tau,\sigma) &=&  \phi^t\cos(\mt\tau)+\widetilde{\phi}^t\sin(\mt\tau)+\widetilde{\psi}^t\sinh(\mt\sigma)+\psi^t\cosh(\mt\sigma)\\&\;\;& + \sum_{\stackrel{n,\omega_n\neq 0}{\omega_n^2=n^2+\mt^2}}c_n\left({\psi}^t_ne^{-i(\omega_n\tau-n\sigma)}+\frac{i}{\mt}(\omega_n-n)\widetilde{\psi}^t_n e^{-i(\omega_n\tau+n\sigma)}\right)\label{allgemein3}\end{eqnarray} with \begin{eqnarray} c_n= \frac{\mt}{\sqrt{ 2\omega_n \left(\omega_n-n\right)  }} .  \end{eqnarray}

\subsection{Dirichlet directions}  In this section we will consider  the bulk fields spanning along a Dirichlet direction with  boundary conditions  \begin{eqnarray} \label{boundarydir1} X^s(\tau,\sigma=0) = y^s_0;\;\;\; X^s(\tau,\sigma=\pi) = y^s_{\pi}\end{eqnarray} and  \begin{eqnarray} \label{boundarydir2} 0= \left( \psi_+^s+\rho\psi_-^s   \right)(\tau, \sigma=0,\pi).\end{eqnarray} Here $\rho=\pm 1$ distinguishes  as usual between the brane / antibrane cases. Our  discussion proceeds  in this part   along the lines of the  $(0,0)-$instanton construction  from \cite{gabgre2}, but differs mildly  in the fermionic sector due to our choice of LG-fermions as discussed in section \ref{landau}.  
\\ From (\ref{allgemein1}),  the boundary conditions (\ref{boundarydir1}) and  (\ref{boundarydir2}) lead to the bosonic mode expansion 
 \begin{eqnarray} \label{dirbos1} X^s (\tau, \sigma)= x^s_0 \cosh(\mt\sigma)+\frac{x^s_\pi-x_0^s\cosh(\mt\pi)}{\sinh(\mt\pi)}\sinh(\mt\sigma)  \nonumber \\ -\sqrt{2} \sum_{n\in\mathbb{Z}\backslash\{ 0\}} \frac{1}{\omega_n}e^{-i\omega_n\tau}a^s_n \sin(n\sigma)\end{eqnarray}  with $\omega_n=\mbox{sgn}(n)\sqrt{n^2+\mt^2}$, compare  for example with \cite{gabgre2}. \\[1ex] For the fermions spanning between a brane-brane configuration we deduce  for $t\in{\cal D}_-$ 
 \begin{eqnarray} \label{dirfer1}\psi^t_+(\tau,\sigma)&=& -\psi^t e^{-\mt \sigma} +\sum_{n\in  \mathbb{Z}\backslash\{ 0\}} c_n \left(\widetilde{\psi}_n^t e^{-i(\omega_n\tau+n\sigma)}-i\frac{\omega_n-n}{\mt}{\psi}_n^t e^{-i(\omega_n\tau-n\sigma)}\right)\\  \label{dirfer2} \psi_-^t(\tau,\sigma)&=& \psi^t e^{-\mt\sigma}+ \sum_{n\in  \mathbb{Z}\backslash\{ 0\}} c_n \left({\psi}_n^t e^{-i(\omega_n\tau-n\sigma)}+i\frac{\omega_n-n}{\mt}\widetilde{\psi}_n^t e^{-i(\omega_n\tau+n\sigma)}\right) \end{eqnarray}
with  the identifications \begin{equation} \widetilde{\psi}_n^t = -\frac{n-i\mt}{\omega_n}\psi^t_n.\end{equation} As explained before, the solutions  along the directions $t\in {\cal D}_+$   are  obtained from (\ref{dirfer1}) and (\ref{dirfer2}) by using $\mt\rightarrow -\mt$. \\[2ex]   The fermionic fields spanning between a brane/antibrane combination have the same structure as presented in  (\ref{dirfer1}) and (\ref{dirfer2}). In this case, however, the zero modes  $\psi^t$  are absent and the nonzero  modings  have to fulfil  either \begin{eqnarray} \label{modingspecial} e^{2\pi i n} = -\frac{n-i\mt}{n+i\mt} \;\;\; \text{or}\;\;\;    e^{2\pi i n} = -\frac{n+i\mt}{n-i\mt};\;\;\;\; n\neq 0\end{eqnarray}  depending on whether $t\in{\cal D}_-$ or $t\in{\cal D}_+$, compare again with the  discussion of the (0,0) instanton in  \cite{gabgre2}.

\subsubsection{Quantisation}\label{dirquantisation} By requiring the standard canonical commutators   as summarised in the appendix \ref{appquantisation} we obtain the  commutation relations for the modes  introduced in the last section to \begin{eqnarray} [a_m^i, a_n^j]& =& \omega_m\delta^{ij}\delta_{m+n}\label{dirmodesbos1}\\ \label{dirmodesfer1} \left\{\psi_m^r, \psi_n^s\right\} &=& \delta^{rs}\delta_{m+n}\\ \left\{\psi^r,\psi^s\right\}&=& \frac{2\pi \mt}{1-e^{-2\pi \mt}}\delta^{rs}= \frac{\pi\mt e^{\pi\mt}}{\sinh(\pi\mt)}\delta^{rs}. \label{dirmodesfer2}\end{eqnarray}
The anticommutators  are   written down for  parameters $r,s$ ranging in  $ {\cal D}_-$. Some details of  the derivations,  in particular of (\ref{dirmodesfer2}),  can be found in the appendix \ref{appquantisation}.

\subsection{Neumann directions}  In this part we will consider the mode expansions for the new Neumann type boundary conditions including   contributions of  the boundary Lagrangian as discussed above. 
We will work with the boundary fields presented in equation  (\ref{sol4}) whose parameters fulfil
 \begin{eqnarray} A^{\ih} B^{\ih} = -e^{i\beta}\mt ;\;\;A^{\ih}D^{\ih}+C^{\ih}B^{\ih}=0 \label{zwischen6}\end{eqnarray} to obey (\ref{susycond2}), ensuring  in particular the conservation of  a ${\cal N}=2$ supersymmetry structure. As before, there is no sum over hatted indices.   Using (\ref{sol4}), the boundary potential $B$  from (\ref{susycond1})  takes on the structure \begin{eqnarray} \label{bintegrable1} B({\bf z}, {\bf \overline{z}})=\sum_{i\in {\cal N}} \left(\widetilde{b}^i z^i \overline{z}{}^{\io} + \widetilde{k}^i \zo^{\io}+\overline{\widetilde{k}}{}^{\io} z^i\right)+\text{const}  \end{eqnarray} by using the convenient combinations  \begin{eqnarray} \widetilde{b}^i=\widetilde{b}^I=\widetilde{b}^{I+4}=\frac{A^{\ih}\overline{A}{}^{\ih}+B^{\ih}\overline{B}{}^{\ih}}{2} \\ \widetilde{k}^i =\frac{C^{\ih}\overline{A}{}^{\ih}+D^{\ih}\overline{B}{}^{\ih}}{2};\;\;    \overline{\widetilde{k}}{}^{i} =\frac{\overline{C}{}^{\ih}{A}^{\ih}+\overline{D}{}^{\ih}{B}^{\ih}}{2}.\end{eqnarray} With   (\ref{zwischen6}) we furthermore have \begin{eqnarray}\label{zwischen5}  A^{\ih}\overline{A}{}^{\ih}= \widetilde{b}^i\pm\sqrt{(\widetilde{b}^{\ih})^2-\mt^2} ;\;\;\; k^i=\pm\frac{C^{\ih}}{A^{\ih}}\sqrt{(\widetilde{b}^{\ih})^2-\mt^2}\end{eqnarray} for  \begin{eqnarray} 0< \mt\leq \widetilde{b}^i\;\;\; \text{and}\;\;\; i\in{\cal N}_-.\end{eqnarray}
In this section we will assume throughout $m<b^i$ and comment on the limiting cases  $b^i=m$ and their  relation in the bosonic sector to   previously  known branes  later on in section \ref{limitcompare}.\\   From (\ref{bcond1})-(\ref{bcond4}),  the boundary conditions at $\sigma=0,\pi$ become  \begin{eqnarray} \label{boundcond5} \partial_\sigma X^{I}= \widetilde{b}^{\hat{I}} X^{\hat{I}}+\widetilde{k}^{{I}}\end{eqnarray} for the bosons
 with $I\in{\cal N}$ .  For the fermionic boundary conditions we use  the boundary equations of motion  (\ref{bcond5}) and (\ref{bcond6}) to eliminate the boundary fermions  from (\ref{bcond3}) and (\ref{bcond4}) and derive  \begin{eqnarray} \label{boundcond7}\pt\left(\psi_+^{I}-\rho\psi_-^{I}\right) &=& (\widetilde{b}^{\Ih}-\rho \mt)\left(\psi_+^{\Ih}+\rho\psi_-^{\Ih}\right)\\ \pt\left(\psi_+^{I+4}-\rho\psi_-^{I+4}\right) &=& (\widetilde{b}^{\Ih}+\rho\mt)\left(\psi_+^{\Ih+4}+\rho\psi_-^{\Ih+4}\right)\label{boundcond6}\end{eqnarray}  for the fermionic bulk fields with $\sigma=0,\pi$ and $I\in {\cal N}_-$. Both cases are formulated in a real basis and  the parameter $\rho$ distinguishes as before between the brane / antibrane boundary conditions. \\[1ex]
 Using the general solution  (\ref{allgemein1}) together  with the boundary conditions (\ref{boundcond5}) the bosonic mode expansions  along the Neumann directions are found  to be  \begin{eqnarray} \nonumber X^I(\tau, \sigma) &=& N^I\cosh(\mt\sigma)+\widetilde{N}^I\sinh(\mt\sigma)  +P^{\Ih} e^{\sqrt{(\widetilde{b}^{\Ih})^2-\mt^2}\tau}\;e^{\widetilde{b}^{\Ih} \sigma}+Q^{\Ih} e^{-\sqrt{(\widetilde{b}^{\Ih})^2-\mt^2}\tau}\; e^{\widetilde{b}^{\Ih}\sigma}\\ &\;& +\frac{i}{\sqrt{2}}\sum_{n\in\mathbb{Z}\backslash\{ 0\}} \frac{1}{\omega_n} \left(a^I_n e^{-i(\omega_n\tau-n\sigma)}+\widetilde{a}^I_n e^{-i(\omega_n\tau+n\sigma)}\right)\label{neubos1}\end{eqnarray} with   
\begin{eqnarray} \widetilde{a}^I_n &=& \frac{n+i\alphat^{\Ih}}{n-i\alphat^{\Ih}}a_n^{\Ih}\end{eqnarray} and 
\begin{eqnarray}\label{zwischenn1}N^I &=& \frac{\widetilde{b}^{\Ih}\cosh\frac{\mt\pi}{2}-\mt\sinh\frac{\mt\pi}{2}}{(\mt^2-(\widetilde{b}^{\Ih})^2)\cosh\frac{\mt\pi}{2}}\widetilde{k}^{\Ih}    \\      \widetilde{N}^I &=& \frac{\mt\cosh\frac{\mt\pi}{2}-\widetilde{b}^{\Ih}\sinh\frac{\mt\pi}{2}}{(\mt^2-(\widetilde{b}^{\Ih})^2)\cosh\frac{\mt\pi}{2}}\widetilde{k}^{\Ih} .  \label{zwischenn2}    \end{eqnarray}
The special modes $P^I, Q^I$ with a time dependency proportional to  $e^{\pm\sqrt{(\widetilde{b}^{I})^2-\mt^2}\tau}$ are of the same type as those appearing in \cite{chu, mattik} in the treatment  of open  strings in the plane wave background  under the inclusion of   a nontrivial  ${\cal F}^{IJ}$-field.  They play a crucial r$\hat{\text{o}}$le  in the  quantisation  to be discussed in the next section.\\
For the fermions spanning between a brane/brane pair with $\rho=1$ we obtain from (\ref{allgemein2}),(\ref{allgemein3}) and the boundary conditions (\ref{boundcond7})  the solutions 
\begin{eqnarray} \nonumber \psi_+^I(\tau,\sigma) &=& -\psi^I e^{-\mt\sigma} +e^{-\sqrt{(\widetilde{b}^{\Ih})^2-\mt^2}\tau} e^{\widetilde{b}^{\Ih}\sigma}\chi^{\Ih}+e^{\sqrt{(\widetilde{b}^{\Ih})^2-\mt^2}\tau} e^{\widetilde{b}^{\Ih}\sigma}\frac{\sqrt{(\widetilde{b}^{\Ih})^2-\mt^2}+\widetilde{b}^{\Ih}}{\mt}\widetilde{\chi}^{\Ih}\\ &\label{neufer1} \;\;&+ \sum_{n\in  \mathbb{Z}\backslash\{ 0\}} c_n \left(\widetilde{\psi}_n^I e^{-i(\omega_n\tau+n\sigma)}-i\frac{\omega_n-n}{m}\psi_n^I e^{-i(\omega_n\tau-n\sigma)}\right)     
\\ \psi_-^I(\tau,\sigma) &=& \psi^Ie^{-m\sigma} +e^{\sqrt{(\widetilde{b}^{\Ih})^2-\mt^2}\tau} e^{\widetilde{b}^{\Ih}\sigma}\widetilde{\chi}^{\Ih}+e^{-\sqrt{(\widetilde{b}^{\Ih})^2-\mt^2}\tau} e^{\widetilde{b}^{\Ih}\sigma}\frac{\sqrt{(\widetilde{b}^{\Ih})^2-\mt^2}+\widetilde{b}^{\Ih}}{\mt}{\chi}^{\Ih}   \nonumber \\ &\;\;&+ \sum_{n\in  \mathbb{Z}\backslash\{ 0\}} c_n \left({\psi}_n^I e^{-i(\omega_n\tau-n\sigma)}+i\frac{\omega_n-n}{m}\widetilde{\psi}_n^I e^{-i(\omega_n\tau+n\sigma)}\right)    \label{neufer2}\end{eqnarray} 
with \begin{equation} \widetilde{\psi}^{I}_n=\frac{\omega_n}{n+i\mt}\frac{n+i\widetilde{b}^{\Ih}}{n-i\widetilde{b}^{\Ih}}\psi^{\Ih}_n\end{equation}  and  $I\in{\cal N}_-$.  The modes $\chi^I$ and $\widetilde{\chi}^I$ correspond  to the bosonic operators  $P^I,Q^I$, compare for example with \cite{mattik}. As described there, the terms in (\ref{neufer1}) and (\ref{neufer2}) containing these special modes  fulfil the conditions (\ref{boundcond7}), (\ref{boundcond6})  for all $\sigma$ and not only on the boundary.\\ The remaining fermionic solutions  along the $I\in {\cal N}_+$ directions are again deduced  by sending  $\mt\rightarrow -\mt$ in (\ref{neufer1}) and (\ref{neufer2}). In particular,  one obtains the mode identifications  for the nonzero modes  in this case to   \begin{equation} \widetilde{\psi}^{I+4}_n=\frac{\omega_n}{n-i\mt}\frac{n+i\widetilde{b}^{\Ih}}{n-i\widetilde{b}^{\Ih}}\psi^{\Ih+4}_n.\end{equation}
As for  the Dirichlet directions, the mode expansion for strings stretching between a brane / antibrane pair deduces  from (\ref{neufer1}) and (\ref{neufer2}) by dropping the zero modes $\psi^I$,  but retaining the special modes $\chi^I$ and $\widetilde{\chi}^I$.  Furthermore, the moding for the nonzero modes again has to fulfil either  
\begin{eqnarray} e^{2\pi i n} = -\frac{n-i\mt}{n+i\mt} \;\;\; \text{or}\;\;\;    e^{2\pi i n} = -\frac{n+i\mt}{n-i\mt}\end{eqnarray} depending on whether $I\in {\cal N}_-$ or $I\in{\cal N}_+$.

\subsubsection{Quantisation} \label{neuquantisation} The standard canonical conditions (\ref{cano1})-(\ref{cano6}) lead in the Neumann case to  the following commutators. For the bosons we obtain 
\begin{eqnarray}\label{neumodesbos1} [P^I, Q^J] &=& \delta^{\Ih J} \frac{2\pi i \alphat^{\Ih}
}{\sqrt{(\alphat^{\Ih})^2-\mt^2}} \frac{1}{1-e^{2\pi\alphat^{\Ih}}}\\ {}  \label{neumodesbos2}  [a^I_m, a^J_n] &=& \omega_m \delta^{IJ}\delta_{m+n}\end{eqnarray}  whereas for the fermions  \begin{eqnarray} \label{neumodesfer1} \left\{\psi^I_m, \psi^J_n\right\} &=&  \delta^{IJ}\delta_{m+n}\\  \label{neumodesfer2} \left\{\psi^I, \psi^J\right\} &=& -\frac{2\pi \mt\delta^{\Ih J}}{1-e^{-2\pi \mt}} \frac{\mt-\alphat^{\Ih}}{\mt+\alphat^{\Ih}}\\  \label{neumodesfer3}\left\{\chi^I,\widetilde{\chi}^J\right\} &=& \frac{2\pi \alphat^{\Ih}\delta^{\Ih J}}{1-e^{2\pi\alphat^{\Ih}}}\;\frac{\sqrt{(\alphat^{\Ih})^2-\mt^2}-\alphat^{\Ih}}{\alphat^{\Ih}+\mt}.\end{eqnarray} The fermionic  relations are  again formulated for $I,J\in {\cal N}_-$ only.  Some  details of the derivations are  presented  in the  appendix \ref{appquantisation}. 
  
\subsection{Boundary fermions} \label{boundaryfermions}
In the last section the  boundary fermionic  fields were  eliminated from the remaining boundary conditions  by using  their equations of motion. In this section we reconsider this situation and  present  the  explicit solution for the boundary fermions  as a suitable combination of fermionic bulk fields evaluated on the boundary. \\   For  our choice of diagonal matrices $F,Q$ all non-diagonal elements of $A, A^\dagger$ in (\ref{boundlagra})  decouple from the remaining fields and   we can  therefore  concentrate   on the diagonal components.   For these  elements  we have to solve the equations of motion (\ref{bcond5}) and (\ref{bcond6}) by using (\ref{sol4}).  For notational simplicity we will write down  only expressions for   fermions corresponding to the $z^1$ direction and suppress  for this case irrelevant indices. \\[1ex]  By using (\ref{boundcond7}) and (\ref{boundcond6}) in the equations of motion (\ref{bcond5}) and (\ref{bcond6}) we obtain  the boundary fermions to  \begin{eqnarray} \label{sola1} a(t)= a_0+\frac{\overline{A}+B}{2(\alphat-\mt)}\left(\psi_+^1-\psi_-^1\right)-i\frac{\overline{A}-B}{2(\alphat+\mt)} \left(\psi_+^5-\psi_-^5\right)
\\ \overline{a}(t)= \overline{a}_0+\frac{{A}+\overline{B}}{2(\alphat-\mt)}\left(\psi_+^1-\psi_-^1\right)+i\frac{{A}-\overline{B}}{2(\alphat+\mt)} \left(\psi_+^5-\psi_-^5\right) \label{sola2}\end{eqnarray} with   constant fermions  $a_0, \overline{a}_0$.  Using    so far only to the differentiated boundary conditions  (\ref{boundcond7}) and (\ref{boundcond6}), we have  to test   whether there are  additional constraints on these extra fermions. From the (undifferentiated)  conditions (\ref{bcond3}) and (\ref{bcond4}) we obtain \begin{equation} 0=B \overline{a}_0+A{a}_0\end{equation} which amounts to $a_0=\overline{a}_0=0$ by using the explicit expressions for $A$ and $B$ from  (\ref{zwischen6}) and (\ref{zwischen5}) with $\alphat>\mt$.
 For our solution (\ref{sol4}) all boundary fermions in (\ref{boundlagra}) therefore either decouple  from the remaining fields or are expressible in terms of bulk functions restricted to the boundary.\\[2ex]  For consistency of the last result, the fermionic anticommutation relations for the bulk fields derived in  section \ref{neuquantisation}  should   reproduce the expected  anticommutators   for the boundary fermionic fields $a(t)$ and $\overline{a}(t)$.  To determine these relations  we have to evaluate expressions like
 \begin{equation} \label{star} (\star)=\left\{\psi_+^1(\tau,\sigma)-\psi_-^1(\tau,\sigma),\psi_+^1(\tau,\overline{\sigma})-\psi_-^1(\tau,\overline{\sigma})\right\}\end{equation} at the boundaries. This  is, different to the bulk, relatively  subtle due to   potential divergencies. Using (\ref{neumodesfer2}) and (\ref{neumodesfer3}) we obtain   \begin{eqnarray} (\star) =-\frac{8\pi \mt}{1-e^{-2\pi \mt}}\frac{\mt-\alphat}{\mt+\alphat} e^{-\mt(\sigma+\overline{\sigma})} -\frac{8\pi \alphat}{1-e^{2\pi\alphat}} \frac{\mt-\alphat}{\mt+\alphat}e^{\alphat(\sigma+\overline{\sigma})} \\ +2\sum_{n\neq 0} \left(e^{in(\sigma-\overline{\sigma})}-\frac{n+i\mt}{n-i\mt}\frac{n-i\alphat}{n+i\alphat}e^{in(\sigma+\overline{\sigma})}\right).\end{eqnarray} After setting  one of the arguments $\sigma, \overline{\sigma}$  equal to the boundary values  $0$ or $\pi$ we have for the infinite sum \begin{eqnarray} 4i(\alphat-\mt)\sum_{n\neq 0} \frac{n e^{in(\sigma+\overline{\sigma})}}{(n-i\mt)(n+i\alphat)}=-4i(\alphat-\mt)\oint_{\cal C} dz\;\frac{e^{iz(\sigma+\overline{\sigma})}}{1-e^{2\pi iz}}\frac{z}{(z-i\mt)(z+i\alphat)}\end{eqnarray} where ${\cal C}$ is a contour running infinitesimally above and below the real axis, compare with \cite{gabgre2}.  By closing the contours the residues cancel out with the first terms in $(\star)$  and we finally obtain   \begin{eqnarray} \left\{\psi_+^1(\tau,\sigma)-\psi_-^1(\tau,\sigma),\psi_+^1(\tau,\overline{\sigma})-\psi_-^1(\tau,\overline{\sigma})\right\}=\pm 4\pi (\alphat -\mt)\\  \left\{\psi_+^5(\tau,\sigma)-\psi_-^5(\tau,\sigma),\psi_+^5(\tau,\overline{\sigma})-\psi_-^5(\tau,\overline{\sigma})\right\}= \pm4\pi (\alphat +\mt)  \end{eqnarray} at $\sigma=\overline{\sigma}=0$ and $\sigma=\overline{\sigma}=\pi$, respectively. 
Using (\ref{sola1}) and (\ref{sola2}) this  leads to  the anticommutators  \begin{eqnarray} \left\{a(t),a(t)\right\}&=&0\\ \left\{\overline{a}(t),\overline{a}(t)\right\} &=& 0 \\ \left\{a(t),\overline{a}(t)\right\}&=&\pm 4\pi   \end{eqnarray} which is  the expected result. The signs originate here in  different overall signs appearing in the boundary Lagrangian (\ref{boundlagra}) at the two boundaries.

\subsection{The ${\cal N}=2$ superalgebra} In this final section  of our open string treatment  we will determine the Hamiltonians of the previously discussed configurations and for the brane-brane situation also the resulting ${\cal N}=2$ supercharges. The expressions for the  Hamiltonians will be put to  use in section \ref{openclose}.\\ The conserved  supercharges are calculated from the equations (\ref{qboundary1}) and (\ref{qboundary2}) established in section  \ref{branesusy2} whereas the Hamiltonians deduce from the following (closed string) conserved fluxes 
\begin{align} \label{fluxham1}    T_2 = g_{j\jo} \left(\partial_+ \zo^{\jo}\partial_+ z^j+\frac{i}{2} \psio_+^{\jo}\stackrel{\leftrightarrow}{\partial}_+\psi_+\right)\;&\;\;  \theta_0  = g_{j\jo} \left( -m^2\zo^{\jo}z^j-\frac{i}{2} \psio^{\jo}_+\stackrel{\leftrightarrow}{\partial}_-\psi_+^j\right)\\ \overline{ T}_2 = g_{j\jo} \left(\partial_- \zo^{\jo}\partial_- z^j+\frac{i}{2} \psio_-^{\jo}\stackrel{\leftrightarrow}{\partial}_-\psi_-\right)\;&\;\; \overline{\theta}_0  = g_{j\jo} \left( -m^2\zo^{\jo}z^j-\frac{i}{2} \psio^{\jo}_-\stackrel{\leftrightarrow}{\partial}_+\psi_-^j\right) \label{fluxham2}\end{align} which fulfil on-shell \begin{eqnarray} \partial_-T_2=\partial_+\theta_0;\;\;\;\; \partial_+\overline{T}_2=\partial_-\overline{\theta}_0.\end{eqnarray}
\subsubsection{Dirichlet directions} 
Using  the fluxes (\ref{fluxham1}) and (\ref{fluxham2}) the open string  Hamiltonian along the Dirichlet directions for a brane/brane configuration  becomes with the mode expansions (\ref{dirbos1})-(\ref{dirfer2}) in  the overall normalisation  explained in detail in \cite{gabgre2} 
\begin{eqnarray} \frac{X^+}{2\pi}H^{\text{open}}&=&\nonumber \frac{\mt}{2\sinh(\mt\pi)}\sum_{a\in {\cal D}}\left(\cosh(\mt\pi)\left({x}^a_0x_0^a+{x}^a_\pi x_\pi^a \right)-2{\ x}^a_0{ x}^a_\pi\right)\\ &\;\;\;\;&+2\pi \sum_{\stackrel{n>0}{a\in{\cal D}}}\left( a^a_{-n}a^a_n+\omega_n \psi^a_{-n}\psi^a_n\right),\label{hamildir}\end{eqnarray}
 where the summation index $a$ is understood to range over all Dirichlet directions.\\[1ex] The Hamiltonian for the brane/antibrane configuration has the same structure with a fermionic nonzero moding as given in (\ref{modingspecial}). In this case there is also  an overall normal ordering constant  which will be implicitly determined in section \ref{openclose}.\\[2ex] The contribution to the overall ${\cal N}=2$ supercharges  for a brane/brane configuration with $\rho=1$ becomes  \begin{eqnarray} \nonumber Q = 2\sum_{a\in{\cal D}_-}\left(\left(\psi^a-i\psi^{a+4}\right)\left(x_0^a+ix_0^{a+4}\right)-\left(e^{-\mt\pi}\psi^{a}-ie^{\mt\pi} \psi^{a+4}\right) \left(x_{\pi}^a+ix_{\pi}^{a+4}\right)\right)  \\ +2\pi\sqrt{2} \sum_{{n\neq 0 \atop a\in{\cal D}_-}} c_n\left[ \left(1-i\frac{\omega_n-n}{\mt}\right)\psi^a_n-i\left(1+i\frac{\omega_n-n}{\mt}\right)\psi^{a+4}_n  \right] \left(a^a_{-n}+ia^{a+4}_{-n}\right)\label{qdir} \end{eqnarray} with the corresponding complex conjugated expression for $Q^\dagger$.

\subsubsection{Neumann directions}
Along the Neumann  directions the fluxes (\ref{fluxham1}) and (\ref{fluxham2}) require the inclusion of  boundary currents    in the open string sector   as discussed in section \ref{branesusy2} for the supercharges and  in the appendix \ref{appintegrability} for the higher spin  currents
of the integrable structure. In the present case,  the local boundary field has  the form  \begin{eqnarray} \Sigma_\pi^{(1)} =2\left(B(z,\zo)+ig_{j\jo}\left(\overline{\theta}_-^{\jo}\theta_+^j-\overline{\theta}_+^{\jo}\theta_-^j\right)\right)\end{eqnarray} and the suitable normalised Hamiltonian becomes for open strings stretching between a brane / brane pair
\begin{align}\nonumber    \frac{X^+}{2\pi} H^{\text{open}}=      H_0+   \sum_{I\in{\cal N}}\left[\frac{\mt^2-(\alphat^I)^2}{\alphat^I}\left(e^{2\pi\alphat^I}-1\right)Q^IP^I\right]+2\pi\sum_{{n> 0 \atop I\in{\cal N}}}\left( a^I_{-n}a^I_n+ \omega_n\psi^I_{-n}\psi^I_n \right) \\  + i  \sum_{I\in{\cal N}_-}\left[(e^{2\pi\alphat^I}-1)\frac{ \left({(\alphat^I)^2-\mt^2}\right)^{\frac{3}{2}}}{\alphat^I \mt^2} \left(\sqrt{(\alphat^I)^2-\mt^2}+\alphat^I\right)\left( \frac{\chi^I\widetilde{\chi}^I}{\alphat^I-\mt}+  \frac{\chi^{I+4}\widetilde{\chi}^{I+4}}{\alphat^I+\mt}\right)\right]     \label{neuham}\end{align} 
with   \begin{eqnarray} \nonumber 2H_0  &=&\sum_{I\in{\cal N}}\left[\left(\mt\cosh(\mt\pi)-\alphat^I\sinh(\mt\pi)\right)\sinh(\mt\pi) \left(N^IN^I+\widetilde{N}^I\widetilde{N}^I\right) \right. \\ \nonumber &\;\;&\;\;\;\;\;\; +2\left(\mt\sinh(\mt\pi)-\alphat^I\cosh(\mt\pi)\right)\sinh(\mt\pi)N^I\widetilde{N}^I\\   &\;\;&\;\;\;\;\;\;  \left. -2\widetilde{k}^I\left(N^I(\cosh(\mt\pi)-1)+\widetilde{N}^I\sinh(\mt\pi)\right)\right]\end{eqnarray}   as contribution from the bosonic zero modes.  Using (\ref{zwischenn1}) and (\ref{zwischenn2}) this simplifies to   \begin{eqnarray}  H_0=\mt \sum_{I\in{\cal N}}\frac{\tanh\frac{\mt\pi}{2}}{(\alphat^I)^2-\mt^2} \widetilde{k}^I \widetilde{k}^I.\end{eqnarray}  The Hamiltonian (\ref{neuham}) is already presented in its normal ordered form by implicitly defining   $P^I$ and $\widetilde{\chi}^I$ as  annihilation operators for the special zero-modes. With these choices the corresponding normal ordering constants cancel.\\[1ex] The Hamiltonian for open strings in between a brane - antibrane pair  also has the structure (\ref{neuham}). In that case, however, the fermionic moding has to fulfil (\ref{modingspecial}) and there also appears a nonzero normal ordering constant. It solely originates from the nonzero modes and takes on  the same value as in the previously discussed  Dirichlet case. \\[2ex]
The contributions to the supercharge in the case of strings in between two branes with $\rho =0$  is finally  obtained to 
  \begin{align} \nonumber Q  =\sum_{I\in {\cal N}_-}&\left[2\left(\widetilde{k}^I+i\widetilde{k}^{I+4}\right)\left( \frac{e^{-\mt\pi}-1}{\alphat^I-\mt}\psi^I-i \frac{e^{\mt\pi}-1}{\alphat^I+\mt}\psi^{I+4}   \right) \right. \\   &+\frac{e^{2\pi \alphat^I}-1}{\mt\alphat^I}\sqrt{(\alphat^I)^2-\mt^2}\left(\sqrt{\alphat^I+\mt}+\sqrt{\alphat^I-\mt}   \right) \nonumber \\  & \times \left( \sqrt{\alphat^I+\mt}\chi^I +i   \sqrt{\alphat^I-\mt}\chi^{I+4}\right) \left(P^I+iP^{I+4}\right) \nonumber   \\ \nonumber &-\frac{e^{2\pi \alphat^I}-1}{\mt\alphat^I}\sqrt{(\alphat^I)^2-\mt^2} \left(\sqrt{\alphat^I+\mt}+\sqrt{\alphat^I-\mt}   \right) \nonumber \\ & \left. \times \left( \sqrt{\alphat^I+\mt} \widetilde{\chi}^I +i \sqrt{\alphat^I-\mt} \widetilde{\chi}^{I+4}\right) \left(Q^I+iQ^{I+4}\right)   \nonumber \right] \\  +2\pi\sqrt{2}& \sum_{{n\neq 0 \atop I\in\cal{N}_-}} \left. c_n\left[ \left(1-i\frac{\omega_n-n}{\mt}\right)\psi^I_n-i\left(1+i\frac{\omega_n-n}{\mt}\right)\psi^{I+4}_n  \right] \left(a^I_{-n}+ia^{I+4}_{-n}\right) \right. \label{qneu} \end{align} with the corresponding complex conjugated expression for $Q^\dagger$.\\[2ex]
 \subsubsection{The superalgebra} 
  Adding up the appropriate  contributions from  (\ref{qdir}) and (\ref{qneu}) corresponding to the particular $(n,n)$-brane under consideration, one obtains the 
 supercharges representing the conserved  
    ${\cal N}=2$ supersymmetry structure of the open string  theory.\\ The anticommutators are found to be 
\begin{eqnarray} \left\{Q, Q^\dagger\right\}& =& \left(8X^+\right)\phantom{\cdot} H^{\text{open}}\\  \left\{Q, Q\right\}& = & 8\pi \mt \sum_{i\in{\cal D}_-} \left(\left(z_0^i\right)^2-\left(z_\pi^i\right)^2\right)\end{eqnarray} which completes our discussion of the open string superalgebra.

 \section{Spacetime supersymmetry and boundary states} \label{spacesusy}
In this section we will study the branes introduced in sections \ref{boundaryintro} and \ref{openstring} from a closed string perspective by formulating them in terms of boundary states. This will on the one hand   confirm our previous results, but is on the other hand  in particular also  suitable for a discussion of  preserved spacetime supersymmetries. As a main result, the spacetime filling $(4,4)$-brane will  be seen to be maximally spacetime supersymmetric. This can be understood in direct analogy to  the other limiting case of the $(0,0)$-instanton.  To have a more straightforward   comparison with  the constructions known for example from \cite{gabgre2},   we will  use a     formulation based on  Green-Schwarz spinors in the closed string channel. 

  \subsection{Gluing conditions}
  By using the standard procedure as for example explained in \cite{ghoshal} or in the context of branes in the plane wave background in \cite{gabgre, gabgre2} one translates the open string boundary  to the corresponding closed string gluing conditions. For the bosonic fields we obtain from (\ref{boundarydir1}) and (\ref{boundcond5})  
  \begin{eqnarray} \label{gluebos1}
0&=&\left.\left(x^r(\tau,\sigma)-y^r_0\right)\right|_{\tau=0}||{\cal B}\rangle\rangle \\ 0&=& \left.\left(\partial_\tau x^I(\tau,\sigma)+i\left(b^{\Ih} x^{\Ih}(\tau,\sigma)+k^I  \right)\right)\right|_{\tau=0}||{\cal B}\rangle\rangle \label{gluebos2}  \end{eqnarray} with $r\in{\cal D}$ and $I\in{\cal N}$.
For the fermions, on the other hand, we have along the Dirichlet directions with $r\in{\cal D}$ \begin{eqnarray}  0&=& \left. \left(   \psi_+^r(\tau,\sigma) -i \rho\psi^r_- (\tau,\sigma)\right)\right|_{\tau =0} ||{\cal B} \rangle\rangle\end{eqnarray} and for the Neumann directions
\begin{eqnarray}     0= \left. \partial_\sigma \left(   \psi_+^I(\tau,\sigma) +i \rho\psi^I_- (\tau,\sigma)\right)+i(b^{\Ih}-\rho m)   \left(   \psi_+^{\Ih}(\tau,\sigma) -i \rho \psi^{\Ih}_- (\tau,\sigma)\right)  \right|_{\tau =0} ||{\cal B} \rangle\rangle \end{eqnarray}    with  $I\in{\cal N}_-$. For $I\in{\cal N}_+$ one  has to interchange  $m\leftrightarrow -m$   and the parameter $\rho=\pm 1$ distinguishes as before between the brane /  antibrane cases.\\ 
Translating these conditions to relations between  Green-Schwarz fermionic fields   by applying the results mentioned in section  \ref{landau}, one derives  the   gluing conditions \begin{align}  0=\eta^*\Gamma^j \left.\left(\widetilde{S}(\tau,\sigma)-i\rho S(\tau, \sigma)\right)\right|_{\tau =0}||{\cal B}\rangle\rangle ;\;\;\;    0=\eta\Gamma^{\jo} \left.\left(\widetilde{S}(\tau,\sigma)-i\rho S(\tau, \sigma)\right)\right|_{\tau =0}||{\cal B}\rangle\rangle \end{align}   
along the Dirichlet directions with $j, \jo\in {\cal D}_-$ and 
 \begin{eqnarray} 0=\eta^*\Gamma^{\hat{j}}\left.\left( \partial_\sigma \left(\widetilde{S}+i\rho S\right)(\tau,\sigma)+i\left(b^{\hat{j}}-m\rho \Pi\right)\left(\widetilde{S}-i\rho S\right) (\tau,\sigma)  \right)\right|_{\tau=0}||{\cal B}\rangle\rangle  \\  0=\eta\Gamma^{\hat{\jo}}\left.\left( \partial_\sigma \left(\widetilde{S}+i\rho S\right)(\tau,\sigma)+i\left(b^{\hat{j}}-m\rho\Pi\right)\left(\widetilde{S}-i\rho S\right) (\tau,\sigma)  \right)\right|_{\tau=0}||{\cal B}\rangle\rangle \end{eqnarray} along the Neumann directions  with $j,\jo\in{\cal N}_-$ and $b^j=b^{\jo}$ as before.
\\ To combine  the fermionic gluing conditions to  a single formula we define  matrices $\Acal, \Bcal$  by the following requirements \begin{eqnarray}\label{ab}   \eta^*\Gamma^i \Acal= \eta^*\Gamma^i;\;\; \eta \Gamma^{\io}\Acal = \eta \Gamma^{\io}\\  \eta^*\Gamma^i \Bcal = b^i \eta^*\Gamma^i;\;\; \eta \Gamma^{\io}\Bcal =b^i \eta \Gamma^{\io}\end{eqnarray} along  the Neumann directions with  $i,\io \in{\cal N}_-$ and \begin{eqnarray} \label{cd} \eta^*\Gamma^r\Acal=\eta^*\Gamma^r\Bcal  =0;\;\; \eta\Gamma^{\overline{r}} \Acal=\eta\Gamma^{\overline{r}} \Bcal=0\end{eqnarray} for the Dirichlet directions with $r,\overline{r}\in{\cal D}_-$. These matrices  especially fulfil 
\begin{eqnarray} \Acal^2=\Acal;\;\;\;\; [\Acal,\Bcal]=[\Acal,\Pi]=[\Bcal,\Pi] =0.\end{eqnarray}
By using $\Acal$ and $\Bcal$ the fermionic gluing conditions  simplify  to the single expression  
\begin{eqnarray} \label{gluefer1} 0=\left.\left(\Acal \partial_\sigma \left(\widetilde{S}+i\rho S\right)(\tau,\sigma)+i\left(\Bcal-m\rho \Pi\right)\left(\widetilde{S}-i\rho S\right) (\tau,\sigma)  \right)\right|_{\tau=0}||{\cal B}\rangle\rangle.\end{eqnarray} 
   
\subsubsection{The  boundary state of the $(n,n)-$brane} \label{theboundarystate}
  By using the closed string mode expansions derived in \cite{metsaev1},   the previously established  field-gluing conditions translate into  relations between   closed string modes acting on the boundary states.  We use the conventions of \cite{gabgre2},  summarised in their appendix A.\\  The bosonic conditions become \begin{eqnarray} \label{gluemodes1} 0= \left(x_0^i-y_0^i\right)||{\cal B}\rangle\rangle;\;\; 0=\left(\alpha_n^i-\widetilde{\alpha}^i_{-n}\right)||{\cal B}\rangle\rangle\end{eqnarray} for $i\in{\cal D}$ and    \begin{eqnarray}\label{gluemodes2} 0= \left(P_0^I+i\left(b^{\Ih} x_0^{\Ih}+k^I\right)\right)||{\cal B}\rangle\rangle;\;\; 0=\left(\alpha_n^I+\frac{\omega_n+b^{\Ih}}{\omega_n-b^{\Ih}}\widetilde{\alpha}_{-n}^{\Ih}\right)||{\cal B}\rangle\rangle   \end{eqnarray}  with $I\in{\cal N}$.   The fermionic gluing conditions translate  into   \begin{eqnarray}\label{gluemodes3} 0&=&\left( \widetilde{S}_0-i\rho S_0\right)||{\cal B}\rangle\rangle\\ \label{gluemodes4}  0&=& \left( \widetilde{S}_n-i\rho \frac{\omega_n-\rho m\Pi}{n} \left(1-\frac{2\omega_n}{\omega_n-\Bcal}\Acal    \right) S_{-n}  \right)||{\cal B}\rangle\rangle.\end{eqnarray}
Finally, by using the following zero mode combinations from \cite{metsaev1, gabgre2} \begin{eqnarray}a^r=\frac{1}{\sqrt{2m}}\left(p_0^r+imx_0^r\right);\;\; \overline{a}^r=\frac{1}{\sqrt{2m}}\left(p_0^r-imx_0^r\right)\end{eqnarray} the bosonic zero mode gluing conditions furthermore  take on the  structure  \begin{eqnarray} 0&=& \left(\overline{a}^i-a^i+i\sqrt{2m}y^i\right)||{\cal B}\rangle\rangle   \label{gluemodes5} \\  \label{gluemodes6}  0&=& \left(\overline{a}^I+\frac{m+b^{\Ih}}{m-b^{\Ih}}a^{\Ih}+i\frac{\sqrt{2m}k^{\Ih}}{m-b^{\Ih}}\right)||{\cal B}\rangle\rangle.\end{eqnarray} After determining the closed string gluing conditions in (\ref{gluemodes1})-(\ref{gluemodes6}) it is now straightforward to write down the corresponding boundary state up to an overall normalisation. This normalisation  ${\cal N}_{(n,n)}$ is obtained  from the results presented in  section \ref{openclose} in the standard procedure by comparing a suitable closed string boundary state overlap with the corresponding open string one loop partition function. As in the instanton case from \cite{gabgre2}, the normalisation ${\cal N}_{(n,n)}$ turns out to  be   \begin{eqnarray} \label{normalisation} {\cal N}_{(n,n)} = (4\pi m)^2\end{eqnarray} up to an irrelevant overall constant phase.  With  the gluing conditions  (\ref{gluemodes1})-(\ref{gluemodes6}), the boundary state takes on  the form 
\begin{eqnarray} \nonumber ||{\cal B}\rangle\rangle = {\cal N}_{(n,n)}\text{exp}\left[ \sum_{r=1}^\infty \sum_{i\in{\cal D}} \frac{1}{\omega_r}\alpha^i_{-r}\widetilde{\alpha}^i_{-r}  +\sum_{r=1}^\infty \sum_{I\in{\cal N}} \frac{1}{\omega_r}\frac{\omega_r+b^I}{\omega_r-b^I}\alpha^I_{-r}\widetilde{\alpha}^I_{-r}  \right. \\ \left. +i \rho\sum_{r=1}^\infty \sum_{a,b}\left(\frac{\omega_r-m\rho\Pi}{r} \left( 1- \frac{2\omega_r}{\omega_r-\Bcal}\Acal\right)   \right)_{ab} \widetilde{S}^a_{-r}S^b_{-r}\right] ||{\cal B}_0\rangle\rangle \label{bstate1}\end{eqnarray} 
  with \begin{eqnarray} \label{bstate2} ||{\cal B}_0\rangle\rangle= \prod_{I\in{\cal N}}\prod_{i\in{\cal D}} \left({\cal B}^I_0 {\cal B}^i_0 \right)|0,\rho\rangle_f\end{eqnarray} and \begin{eqnarray}\label{bstate3} {\cal B}_0^i &=& \text{exp}\left[ \left(\frac{1}{2} a^{\ih}a^{\ih}-i\sqrt{2m}y^{\ih}a^{\ih}\right)\right]  e^{-\frac{m}{2}y^{\ih}y^{\ih}}  \\  {\cal B}_0^I  &=&\exp \left[-\left( \frac{1}{2} \frac{m+b^{\Ih}}{m-b^{\Ih}}a^{\Ih}a^{\Ih}+i\frac{\sqrt{2m}k^{\Ih}}{m-b^{\Ih}}a^{\Ih}   \right)  \right] e^{-\frac{m}{2}\frac{k^{\Ih}k^{\Ih}}{(b^{\Ih})^2-m^2}}.\label{bstate4}\end{eqnarray}
  The fermionic vacuum state $|0,\rho\rangle_f$  is finally determined by the condition (\ref{gluemodes3}), compare for example with \cite{gabgre2}.

 \subsection{Spacetime supersymmetry} \label{spacetimesusy}
 In this section we will determine the preserved (spacetime) supersymmetries  of the boundary state (\ref{bstate1}). Our discussion from the open string point of view in section \ref{openstring} together with the considerations from section \ref{landau} ensures  at least two preserved supersymmetries on (\ref{bstate1}). Under certain conditions, however, some  $(n,n)$- branes preserve additional  supercharges.  A certain  class of $(4,4)$-branes, for example,  will be seen to be even maximally supersymmetric.\\[1ex] 
 In the conventions of \cite{gabgre2} the (dynamical) supersymmetries of the plane wave background take on the form \begin{eqnarray}  \sqrt{2P^+} Q= \sum_r\left[p_0^r\gamma^rS_0-mx_0^r\gamma^r\Pi \widetilde{S}+\sum_{n\neq 0} c_n\left(\gamma^r \alpha_{-n}^rS_n+i\frac{\omega_n-n}{m}\gamma^r\Pi \widetilde{\alpha}_{-n}^r\widetilde{S}_n  \right) \right] \\  \sqrt{2P^+} \widetilde{Q}= \sum_r\left[p_0^r\gamma^r\widetilde{S}_0+mx_0^r\gamma^r\Pi {S}+\sum_{n\neq 0} c_n\left(\gamma^r \widetilde{\alpha}_{-n}^r\widetilde{S}_n-i\frac{\omega_n-n}{m}\gamma^r\Pi {\alpha}_{-n}^r{S}_n  \right)\right]  \end{eqnarray} and the conservation of supersymmetries by the boundary state (\ref{bstate1}) is expressed by  \begin{eqnarray} \label{conservp} 0= P\left(Q+i\rho M\widetilde{Q}\right)||{\cal B}\rangle\rangle\end{eqnarray} with a constant $SO(8)-$spinor matrix $M$ and a suitable projector $P$ whose (maximal)  rank equals the number of preserved (dynamical) supersymmetries.\\[1ex] By using (\ref{gluemodes1}) - (\ref{gluemodes4}) we  derive conditions for the matrices $M$ and $P$ as follows.  From the zero modes along either the Dirichlet or Neumann directions from (\ref{gluemodes1}) and (\ref{gluemodes2}) we obtain \begin{eqnarray} 0=P\left(1-M\right)\;\;\;\Leftrightarrow \;\;\; PM=P \label{suscond1}.\end{eqnarray}
 From the nonzero modes along the  Dirichlet directions from (\ref{gluemodes1}) we have  \begin{eqnarray} 0= P \gamma^i \left( \left(1+\rho\frac{\omega_n-n}{m}\Pi\right)S_n+i \rho\left(1-\rho\frac{\omega_n-n}{m}\Pi\right) \widetilde{S}_{-n}  \right)||{\cal B}\rangle\rangle \end{eqnarray}  and with (\ref{gluemodes4}) \begin{eqnarray} 0= P\gamma^i\Acal \label{suscond2}. \end{eqnarray} 
 From the Neumann directions with the gluing conditions (\ref{gluemodes2}) one furthermore derives  
  \begin{eqnarray} 0= P \gamma^{\Ih} \left(\frac{\omega_n+b^{\Ih}}{\omega_n-b^{\Ih}} \left(1-\rho\frac{\omega_n-n}{m}\Pi\right)S_{-n}-i\rho \left(1+\rho\frac{\omega_n-n}{m}\Pi\right) \widetilde{S}_{n}  \right)||{\cal B}\rangle\rangle \end{eqnarray} from which \begin{eqnarray} 0= P\gamma^{\Ih}   \left[ \omega_n\left(1-\Acal\right)+\Acal\left(b^{\Ih}-\Bcal\right) \right]\label{suszwischen} \end{eqnarray} results by using (\ref{gluemodes4}). As (\ref{suszwischen}) is required to hold  for all $n$ the conditions for preserved supersymmetries finally become
  \begin{eqnarray}  \nonumber 0&=&P\gamma^I\left(1-\Acal\right);\;\;\;\;0=P\gamma^{\Ih}\left(b^{\Ih}-\Bcal\right)\\  0&=& P\gamma^i \Acal  \label{suscond5}\end{eqnarray} with $I\in{\cal N}$ and $i\in{\cal D}$.\\[2ex]
 From (\ref{suscond5}) we can read off the number of conserved supersymmetries for  $(n,n)$-branes with the present boundary conditions. To start with, for $n=0$ one obtains the $(0,0)$-instanton from \cite{skentay2, gabgre2}. It has only Dirichlet directions and from  (\ref{cd}) we furthermore have $\Acal=0$,  that is, $P$ is of maximal rank, implying  a maximally  supersymmetric brane. This is of course exactly the result of \cite{skentay2, gabgre2}.  \\[1ex] The remaining branes preserve at least the ${\cal N}=2$ supersymmetry structure discussed in section \ref{openstring} from an open string point of view. Here this subalgebra is obtained by the projector \begin{eqnarray} P= |\eta\rangle \langle \eta^*|+|\eta^*\rangle\langle\eta |,\end{eqnarray} using  the constant spinor $\eta$  defined in section \ref{landau}.  For the $(1,1)$ and $(3,3)$-branes and in case of pairwise different $b^i$ for the $(2,2)$ and $(4,4)$-branes these exhaust the    conserved supersymmetries.\\[2ex]  For homogenous boundary conditions along the Neumann directions, that is, by  using the same parameter $b$ for all Neumann blocks,  there, however,  appear  additional supersymmetries for the $(2,2)$ and the  $(4,4)$ brane beyond the ${\cal N}=2$ subalgebra. Using the matrices $\Acal, \Bcal$ the situation of homogenous boundary conditions translates into  \begin{eqnarray} \label{mehr1} \Bcal=b \Acal, \end{eqnarray} simplifying (\ref{suscond5}) accordingly.  Evaluating     these conditions with (\ref{mehr1}),  the $(2,2)$ brane is found to be  quarter  supersymmetric, that is,  it  preserves    $4$ supersymmetries and   the $(4,4)$  
brane with $\Acal =1$  and no Dirichlet directions along the transverse coordinates  becomes  finally  even  maximally supersymmetric.\\  Using the classification of \cite{gabgre2}, see also \cite{skentay2}, the $(n,n)$-branes all  belong to the class II branes. Our
 $(4,4)$ therefore adds a maximally supersymmetric brane to this family,  containing  so far only the other extremal case of the $(0,0)$ instanton  and the $(4,0)$, $(0,4)$ branes as half supersymmetric branes.

 \subsection{Boundary conditions with  longitudinal  flux ${\cal F}_{I+}$ } \label{transverseflux}
 In this section we will briefly discuss how to realise deformed Neumann boundary conditions as in (\ref{boundcond5}) by switching on a nonzero  flux ${\cal F}_{I+}$. In the context of plane wave physics  this has been first  discussed in \cite{skentay1} and later on applied in particular in \cite{skentay2, gabgre2, cha1}.\\  In the presence of  a  boundary condensate Neumann conditions read \begin{eqnarray} \partial_\sigma X^r={\cal F}^r_{\;s}\partial_\tau X^s\end{eqnarray} at $\sigma=0,\pi$.  By switching on only  particular longitudinal  components of ${\cal F}$ one obtains \begin{eqnarray} \label{transversef} \partial_\sigma x^r ={\cal F}^r_{\; +} \partial_\tau X^+ \sim {\cal F}^r_{\; +
 } P^+\end{eqnarray} by using the standard lightcone gauge condition on $X^+$. Choosing the  flux ${\cal F}_{I+}$ as a  general affine  function in $X^I$ with appropriate constant factors one obtains  from (\ref{transversef})   the boundary conditions  (\ref{boundcond5}), compare again with \cite{skentay1}.\\[1ex]
 For boundary fields $F, G$ fulfilling the requirements for ${\cal N}=2$ supersymmetry and integrability the boundary conditions (\ref{bcond1}) and (\ref{bcond2}) were seen in section \ref{boundaryintro} to be  independent of the fermionic fields and take on as shown above the standard form for Neumann boundary conditions in the presence of a particular  boundary condensate. Nevertheless, the fermionic boundary conditions  (\ref{bcond3}), (\ref{bcond4}) respectively  (\ref{boundcond7}) and (\ref{boundcond6}) differ clearly from the  conditions usually employed for the fermionic fields in the presence  boundaries. It would be very interesting to  obtain  a  deeper understanding of these conditions and their relation to the  flux ${\cal F}_{I+}$ from (\ref{transversef}) for example by considerations along the lines of \cite{howe1}.\\[2ex] Before discussing  the open/closed duality in section \ref{openclose},  we use (\ref{transversef}) to explain the relation between the open string quantities $\alphat, \mt$ and $\widetilde{k}$ and their  closed string relatives  $b, m, k$.  As discussed  in \cite{gabgre, gabgre2, skentay3} to which  we refer for  a detailed treatment,  one needs to apply different lightcone gauge conditions in the open respectively closed string sectors to deal with branes of the same structure in both cases.  In (\ref{transversef}) this  effectively amounts to interchange the roles of $P^+$ and the lightcone separation $X^+$ of the branes under consideration. As discussed in \cite{gabgre} it follows immediately from this observation that  $m, b, k$ are related to the corresponding open string quantities by  \begin{eqnarray} \label{relation12}  \mt= mt; \;\; \alphat=b t;\;\; \widetilde{k}=k t\end{eqnarray} with \begin{eqnarray} t = \frac{X^+}{2\pi P^+}.\end{eqnarray} The number  $t$ is the modular parameter to appear in section \ref{openclose} where also the relations (\ref{relation12}) will be put to use. 
 
 \subsection{The $b \rightarrow m$ limit.} \label{limitcompare}
To discuss  the limiting situation of  $b=m$  excluded in the previous  discussion   we briefly reconsider    the local boundary field $\Sigma_\sigma(\tau)$  introduced in section \ref{boundaryintro}.  This will especially also establish the maximal supersymmetry of the $(4,4)$ brane in the  open string sector   which so far has been done only for the particular ${\cal N}=2$ subalgebra discussed in section \ref{landau}.  \\ 
From the supercurrents derived in \cite{metsaev1}, used here in the conventions of \cite{gabgre2},  the condition for conserved spacetime supercharges  in the open sector  corresponding to (\ref{conservp})   is given by   \begin{eqnarray}\label{sigmaspace} \partial_\tau \Sigma_\sigma= P \left[  \left(\partial_- x^a \gamma^a S+\mt x^a\gamma^a\Pi \widetilde{S}\right) +{M} \left(-\partial_+x^a\gamma^a\widetilde{S}+\mt x^a\gamma^a\Pi S\right) \right],  \end{eqnarray}  following as before \cite{ghoshal}.  The equation (\ref{sigmaspace}) is again understood to be evaluated at the boundaries $\sigma=0,\pi$ and for the case of the $(4,4)$ brane  to which we restrict attention here one furthermore has  $P=M=1$.\\   By using the bosonic boundary conditions (\ref{boundcond5}) and \begin{eqnarray}  \label{boundaryspec1} 0=\left.\left( \partial_\tau \left(\widetilde{S}-S\right) - \left(\alphat-\mt \Pi\right) \left(\widetilde{S}+S\right)   \right) \right|_{\sigma=0,\pi} \end{eqnarray}  corresponding  to (\ref{gluefer1}),  we derive the following local boundary field \begin{eqnarray}  \Sigma_\pi(\tau) =\sum_I\left[\left( X^I +\frac{\widetilde{k}^I}{\alphat-\mt\Pi}\right)\gamma^I\left(S-\widetilde{S}\right)\right]_{\sigma=\pi}. \label{sigmaspace2}\end{eqnarray}  As it fulfills (\ref{sigmaspace}), the open string theory for the $(4,4)$ brane preserves  the maximal supersymmetry  as expected from the boundary state treatment.\\ From (\ref{sigmaspace2}) it is furthermore 
apparent that the $(4,4)$ remains maximally supersymmetric in the $\alphat\rightarrow \mt$ limit in case of $\widetilde{k}=0$ corresponding to the choice   $C^i=0$ in (\ref{zwischen5}). \\[2ex] It is worth pointing out that  the  bosonic boundary conditions (\ref{boundcond5}) take on in this limit the structure used in \cite{cha1} in an alternative construction of $(n,n)$-branes.  There the authors show from an open string point of view that the common fermionic boundary conditions \begin{eqnarray} 0= \left.\left(\widetilde{S}-MS\right)\right|_{\sigma=0, \pi}\end{eqnarray} with a matrix $M$ as defined in section \ref{braneintro} together with the bosonic boundary conditions  \begin{eqnarray} \partial_\sigma X^I =\pm mX^I;\;\; \partial_\sigma X^{I+4}=\mp m X^{I+4}\end{eqnarray} with $I\in {\cal N}_-$ lead to  $(n,n)-$branes $(n=1,\dots 4)$ which  preserve 4 spacetime supersymmetries. This is expressed by the projectors  
 \begin{eqnarray} P= \frac{1\pm M\Pi}{2}\end{eqnarray} in the conditions  (\ref{sigmaspace}) and  (\ref{conservp}).

 \section{Open-Closed duality}
 \label{openclose}
In this section we consider an important consistency check for the $(n,n)$-boundary states  constructed in section \ref{spacesusy} by testing  the  equality of  the closed string boundary state overlap \begin{eqnarray} \label{overaa} {\cal A}(t) = \langle\langle  {\bf b, k, \bf y_2}|| e^{-2\pi t H^{\text{closed}}P^+} || {\bf b, k, y_1}\rangle\rangle \end{eqnarray} and the one loop open string partition function  \begin{eqnarray}\label{traceaa}  Z(\widetilde{t}) = \text{Tr}\left[ e^{-\frac{X^+}{2\pi}H^{\text{open}} \widetilde{t}}\right].\end{eqnarray} 
The trace in (\ref{traceaa}) runs  over  the states of an open string spanning between  branes with boundary conditions corresponding to the  boundary states in (\ref{overaa}). In the context of plane wave physics this consistency check was first  considered in \cite{gabgre, gabgre2} to which we refer for a detailed discussion. Here we only note that the modular parameters are related by \begin{eqnarray} \widetilde{t} = \frac{1}{t}\end{eqnarray} and the field parameters $b^i, k^i, m$   translate as discussed in section \ref{transverseflux}.\\  We will express  (\ref{overaa}) and (\ref{traceaa}) in terms of special functions defined in \cite{gabgre, gabgre2} as  $m$-dependent  deformations  of the $f$-functions defined in \cite{polchinski} by Polchinski and Cai. \\[2ex]
For open strings spanning between two $(n,n)$-branes of the same type there are fermionic zero modes commuting with the corresponding open string Hamiltonian. As explained for example in \cite{gabgre} these modes lead to vanishing open string partition functions. In the closed string sector this result is confirmed by considering the zero mode part overlap which is also found to vanish, see again \cite{gsw1, gabgre, gabgre2}.\\[1ex] To obtain a nontrivial behaviour we consider the situation of a brane-antibrane configuration.  From (\ref{traceaa}) we have for the open string partition function  along each complex  pair of Dirichlet directions  \begin{eqnarray}\label{z1} Z_{x^i, x^{i+4}}(\widetilde{t})= e^{-\frac{\widetilde{t}\mt}{2 \sinh(\mt \pi) }\sum\limits_{j=i, i+4}\left( \cosh(\mt\pi) \left(y^j_2y_2^j+y_1^jy_1^j\right)-2 y_2^j y_1^j\right)} \frac{  \widehat{g}^{(\mt)}_{4}(\widetilde{q})}{\left(f_1^{(\mt)}(\widetilde{q})\right)^2} \end{eqnarray} with $\widetilde{q}=e^{-2\pi \widetilde{t}}$. For a pair of Neumann directions we deduce analogously  \begin{eqnarray} \label{z2} Z_{x^I, x^{I+4}}(\widetilde{t})=e^{ -\mt\widetilde{t}\sum\limits_{J=I, I+4} \frac{\tanh \frac{\mt\pi}{2}}{(\alphat^J)^2-\mt^2} \widetilde{k}^J\widetilde{k}^J } \frac{  \widehat{g}^{(\mt)}_{4}(\widetilde{q})}{\left(f_1^{(\mt)}(\widetilde{q})\right)^2}.\end{eqnarray}  For the boundary state overlap (\ref{overaa}) one derives 
\begin{align}\label{a1} {\cal A}_{x^i, x^{i+4}}(t) = \exp\left[   -\sum\limits_{j=i, i+4}\left(\frac{m\left(1+q^m\right)\left(y_1^jy^j_1+y_2^jy^j_2\right)}{2\left(1-q^m\right)}+\frac{2m q^{\frac{m}{2}}y_1^jy_2^j}{1-q^m}\right) \right] \frac{g_2^{(m)}(q)}{\left(f_1^{(m)}(q)\right)^2}\end{align} along each pair of Dirichlet directions and \begin{eqnarray} {\cal A}_{x^I, x^{I+4}}(t) =  \exp\left[ -\sum\limits_{J=I, I+4}\frac{mk^Jk^J}{(b^J)^2-m^2} \frac{1-q^{\frac{m}{2}}}{1+q^{\frac{m}{2}}}\right]  \frac{g_2^{(m)}(q)}{\left(f_1^{(m)}(q)\right)^2} \label{a2} \end{eqnarray} 
along a pair of Neumann directions  by using in both cases the normalisation (\ref{normalisation}). The zero mode prefactors in    (\ref{a1}), (\ref{a2})  are for example  calculated by inserting a complete set of coherent states as explained in \cite{gsw1}. \\ From  the  modular transformations properties \begin{eqnarray}f_1^{(m)}(q)= f_1^{(\mt)}(\widetilde{q});\;\;\; g_2^{(m)}(q) =\hat{g}_4^{(\mt)}(\widetilde{q}) \end{eqnarray}  derived in \cite{gabgre, gabgre2},  the open string partition functions (\ref{z1}) and (\ref{z2}) are seen to be equal to the corresponding closed string boundary state overlaps (\ref{a1}) and (\ref{a2}). By this,  the $(n,n)$-branes pass this important consistency check.

\section{Conclusions}
Starting with a boundary Lagrangian containing   fermionic boundary excitations   defined  in analogy to the settings in   \cite{ghoshal, warner1, nepo2} and \cite{ kapustin2,  brunner1} from the  context  of integrable boundary field theories  and  matrix factorisations in string theory, we have constructed   new  integrable and supersymmetric branes in the plane wave background of type $(n,n)$.  As a main result,  the limiting case of the spacetime filling $(4,4)$-brane was  shown to be maximally supersymmetric. This is  in analogy to the other extremal case of the $(0,0)$-instanton from \cite{skentay2, gabgre2}.  \\ The new branes were    constructed in the open and closed string picture,  leading to consistent results in both sectors.  The branes   pass  in particular the open/close-duality check of the equality of  open-string one loop partition functions and corresponding boundary state overlaps, compare with \cite{gabgre, gabgre2}.\\   Whereas the deformed bosonic boundary conditions along the Neumann directions can be understood as a coupling to a nonzero flux ${\cal F}_{+I}$, a statement also supported by the correct reproduction of the relation between gauge-dependent field parameters in the open and closed sector as implied by duality,  the situation for the fermionic sector  is less clear.  It was  demonstrated   that for integrable branes the boundary fermions are consistently determined  by the bulk fields restricted to the boundaries.  However, a more geometric understanding of the resulting  deformed boundary conditions in the fermionic sector,  for example along the lines of \cite{berkovits2, howe1},  remains desirable.

\section*{Acknowledgments}I am grateful to   Matthias Gaberdiel
for encouragement and support and I furthermore thank  Stefan
Fredenhagen and Peter Kaste for discussions. This
research is partially  supported by the Swiss National Science
Foundation and the European Network `Forces-Universe' (MRTN-CT-2004-005104).

\begin{appendix}

\section{LG vs GS spinors} \label{eta}
In this Appendix  we supply some additional details about the identifications between Landau-Ginzburg and Green-Schwarz fermions as briefly discussed in  section \ref{landau}. \\ The identifications (\ref{identi1}), (\ref{ident2}) or the inverted expressions 
\begin{eqnarray}\label{ident3} \psi^i_- = \frac{1}{2} \eta^{*a}\Gamma^i_{ab}S^b\;\;\; &\;&\;\;\; \overline{\psi}^{\io}_- = \frac{1}{2} \eta^{a}\Gamma^{\io}_{ab}S^b\\ \label{ident4}  \psi^i_+ = \frac{1}{2} \eta^{*a}\Gamma^i_{ab}\widetilde{S}^b\;\;\; &\;& \;\;\; \overline{\psi}^{\io}_+ = \frac{1}{2} \eta^{a}\Gamma^{\io}_{ab}\widetilde{S}^b  \end{eqnarray}  can be geometrically interpreted as follows \cite{maldacena}. 
 The  choice of a complex structure in the definition of the Landau-Ginzburg Lagrangian (\ref{lagragen})  figures  out  a $SU(4)$ subgroup of the  $SO(8)$ in whose spinor representations the standard   Green-Schwarz spinors  reside.  Under this subgroup these representations  decompose into \begin{eqnarray} {\bf 8}_-\rightarrow {\bf 4}+ \overline{{\bf 4}}\end{eqnarray} and the summands   correspond to the  spinor fields in (\ref{lagragen}) carrying a vector index.\\ As the superpotential (\ref{superpot1})  already breaks the $SO(8)$-background symmetry  present in flat space down to $SO(4)\times SO(4)\times \mathbb{Z}_2$, the complex structure used in the previous argument actually picks  out  the diagonal $SO(4)$ subgroup of this product.  For this reduced symmetry group   the  fields $\psi^i_\pm,  \overline{\psi}_\pm^{\io}$ transform in the same representation,  explaining  the  seemingly  strange  index structure of   the equations of motion (\ref{eomferlg}).\\[1ex]
Before discussing the ${\cal N}=(2,2)$ worldsheet supersymmetry, we briefly establish  the existence of the spinor $\eta$ with the requirements 
  (\ref{eta1}). Using the properties of the complex Dirac matrices $\Gamma^i, \Gamma^{\io}$,  the spinor   $\eta$ is  immediately  determined to 
  \begin{equation} \label{eta4}\eta= \Gamma_{\oe}\Gamma_{\oz}\Gamma_{\overline{3}}\Gamma_{\overline{4}}\left(\bbbone-\Pi\right) \zeta\end{equation} with a constant real spinor $\zeta=\zeta^*$ of appropriate norm. For example  by employing the explicit  spinor   representation presented in chapter 5 of \cite{gsw1} one can show that the matrix $\Gamma_{\overline{1}}\Gamma_{\oz}\Gamma_{\overline{3}}\Gamma_{\overline{4}}\left(\bbbone-\Pi\right) $ is of \emph{real} rank one, that is, $\eta$ is actually unique up to a sign. Finally, by using \begin{equation} \Pi=\gamma^1\gamma^2\gamma^2\gamma^4=\prod_{i=1}^4\frac{\Gamma^i+\Gamma^{\io}}{\sqrt{2}}\end{equation} the condition $\eta^*=-\Pi \eta$  becomes  \begin{equation} \eta^*=-\frac{1}{4}\Gamma^{\overline{1}}\Gamma^{\oz}\Gamma^{\od}\Gamma^{\ov}\eta=-\frac{1}{4}\Gamma_1\Gamma_2\Gamma_3\Gamma_4\eta.\end{equation}  The lowest weight $su(4)$ state $\eta$ is therefore essentially   related to the corresponding highest weight state by complex conjugation. 
  
  \subsection{${\cal N} = (2,2) $ supersymmetry} \label{n22susy}
In the following we will derive  the relations (\ref{q1}), (\ref{q2})  between the  ${\cal N}=(2,2)$  worldsheet supersymmetry  and the spacetime supercharges from the Green-Schwarz formulation.  This will in particular also lead to an explicit confirmation of the related group theoretical discussion in \cite{gabgre2}. 
\\  The supercurrents for the plane-wave Landau-Ginzburg model described by  (\ref{lagragen}) with superpotential (\ref{superpot1}) are given by \cite{vafa2}  \begin{eqnarray} \label{currents1}G^0_{\pm}=g_{i\io}\partial_\pm \overline{z}^{\io}\psi^i_\pm\pm m\overline{\psi}^{\io}_\mp\overline{z}^{\io}\;\;\; &\;&\;\;\;  G^1_\pm =\mp g_{i\io}\partial_\pm \overline{z}^{\io}\psi^i_\pm + m\overline{\psi}^{\io}_\mp\overline{z}^{\io}\\ \label{currents2}  \overline{G}^0_\pm = g_{i\io} \partial_\pm z^i \overline{\psi}_\pm^{\io}\pm m z^i \psi^i_\mp \;\;\; &\;&\;\;\;  \overline{G}^1_\pm = \mp  g_{i\io} \partial_\pm z^i \overline{\psi}_\pm^{\io}+ m z^i \psi^i_\mp\end{eqnarray} and  lead to the conserved charges  
  \begin{eqnarray} Q_{\pm} &=& \frac{1}{2\pi} \int_0^{2\pi} d\sigma \left(g_{i\jo} \partial_{\pm} \zo^{\jo} \psi_\pm^i \pm m \psio^{\jo}_{\mp}\zo^{\jo} \right) \\
\overline{Q}_{\pm} &=& \frac{1}{2\pi} \int_0^{2\pi}d\sigma \left(g_{\io j} \overline{\psi}^{\io}_{\pm} \partial_{\pm}z^j\pm m \psi^{i}_{\mp}z^i \right)
 \end{eqnarray} representing the ${\cal N}=(2,2)$ worldsheet supersymmetry.  Using the identifications (\ref{identi1}), (\ref{ident2}) and  \begin{eqnarray}  \gamma^r \eta = -i \gamma^{r+4} \eta;\;\;  \gamma^r \eta^*&=&i \gamma^{r+4}\eta^* \end{eqnarray}  from (\ref{eta1}),  we for example deduce\begin{eqnarray}Q_+  &=& \frac{1}{4\pi} \int^{2\pi}_0 d\sigma \left(g_{i\jo} \partial_{+} \zo^{\jo} \left(\eta^*\Gamma^i\widetilde{S}\right) + m \zo^{\jo}\left(\eta \Gamma^{\jo} S\right) \right)\\ &=& \frac{\eta*}{2\pi} \int^{2\pi}_0 d\sigma \left( \partial_+ x^I \gamma^I \widetilde{S}+m x^I\gamma^I\Pi S\right).\end{eqnarray} Comparing this with the expressions for the dynamical spacetime supercharges derived in  \cite{metsaev1}, used here in the conventions of \cite{gabgre2}, we deduce \begin{equation} \frac{Q_+}{\sqrt{2p^+}} = \eta^* \widetilde{Q} = \left(\eta^*\right)_{\dot\alpha} \widetilde{Q}_{\dot\alpha},\end{equation} implicitly using the negative  $SO(8)$ chiralities of the spinors $S,\widetilde{S}$.  In a similar way one  expresses  the remaining  supersymmetries as linear combinations of the spacetime charges  as given in (\ref{q1}), (\ref{q2}).
\\ To relate this result to   the discussion of section $5$ in  \cite{gabgre2}, we only have to note that the condition (\ref{eta1}) requires $\eta$ to be the   bottom state discussed in \cite{gabgre2},     whereas  $\eta^*$ is the corresponding top-state as established beforehand at the end of the last section.

\section{Integrability}\label{appintegrability}
In this appendix we present some additional information about the integrable structure underlying the plane wave theory. For the massive Ising model the  higher spin currents responsible for  integrability  were written down in \cite{zamo1} and are given by  \begin{eqnarray} \nonumber T^f_{n+1} = g_{i\io} \: \psio^{\io}_+\partial_+^n \psi_+^i   ;\;\;\;\; \theta^b_{n-1} = -g_{i\io}\; \psi^i_-\partial_+^n \psio_-^{\io}\\ \overline{T}^f_{n+1} = g_{i\io} \: \psio^{\io}_-\partial_-^n \psi_-^i;\;\;\; \label{currentfer} \overline{\theta}^b_{n-1} = -g_{i\io}\; \psi^i_+\partial_-^n \psio_+^{\io}\end{eqnarray} by concentrating on the for a theory defined on   $S^1\times\mathbb{R}$  relevant  cases. \\  The corresponding bosonic currents are found to equal  \begin{eqnarray} \nonumber T^b_{2n} = g_{i\io}\; \partial_+^n \zo^{\io}\;\partial_+^n z^i;\;\;\;\; \theta^b_{2n-2} = -m^2 g_{i\io} \;\partial_+^{n-1}\zo^{\io} \partial_+^{n-1}z^i\\ \overline{T}^b_{2n} = g_{i\io}\; \partial_-^n \zo^{\io}\;\partial_-^n z^i;\;\;\;\; \overline{\theta}^b_{2n-2} = -m^2 g_{i\io} \;\partial_-^{n-1}\zo^{\io} \partial_-^{n-1}z^i \label{currentsbos}\end{eqnarray}  and the integrable currents for the plane wave theory are given by a suitable combination of (\ref{currentfer}) and (\ref{currentsbos}). Appearing  relative prefactors might for example be  determined by requiring the  cancellation of  separate normal ordering constants in (\ref{currentfer}) and (\ref{currentsbos}) in the quantum theory.\\[1ex]  Treating a free theory, there are nevertheless many different  fluxes like (\ref{currentsbos}). They can for example be obtained  by  taking  the   parts along  single real directions  in  (\ref{currentsbos}) and recombining them in  various ways. This   leads to additional  conserved higher spin  bulk currents, but most choices are  incompatible with the complex structure chosen in the Lagrangian (\ref{lagragen}).\\  Our decision  to consider  the special  combinations (\ref{influx1})-(\ref{intflux4}) is especially  based on the observation that these currents appear as limits of the highly nontrivial higher spin currents of the ${\cal N}=2$ supersymmetric sine-Gordon model. \\  The first nontrivial higher spin currents for this theory were formulated in \cite{kobayashi, nepo2}.  In the language of a Landau-Ginzburg model with superpotential \begin{equation} W=-2ig \cos z+\text{const} \label{wzwischen1} \end{equation} they can be found in \cite{mattik2}.  Reintroducing the standard parameter $\omega$ and rescaling  the coupling constant to $g\rightarrow -\frac{m}{\omega^2}$,  the plane wave like theory with superpotential  $W=im z^2$ is obtained from (\ref{wzwischen1})  in the $\omega\rightarrow 0$ limit.\\ Using the higher spin currents as presented in \cite{mattik2} we obtain furthermore for the first higher spin flux \begin{eqnarray} \label{z23} \frac{T_4}{\omega^2} &=& 2\left( \partial_+^2\zo\;\partial_+^2 z+i\partial_+\psio_+\;\partial_+^2\psi_+\right)+o(\omega^2)\\ \frac{\theta_2}{\omega^2} &=& 2\left(-m^2 \partial_+\zo\;\partial_+ z -im^2 \psio_+\;\partial_+\psi_+\right)+o(\omega^2) \label{z24}.\end{eqnarray}  The formulas  presented in (\ref{influx1}) and (\ref{inftlux2})  and correspondingly in (\ref{intflux3}) and (\ref{intflux4})  differ from (\ref{z23}), (\ref{z24}) only in total derivative terms  included to obtain manifestly real expressions.\\[1ex]  In the boundary theory the currents (\ref{influx1})-(\ref{intflux4}) give rise to the conserved charge \begin{equation}\label{spin3b} I_3=\int_0^\pi d\sigma\; \left(T_4+\overline{T}_4-\theta_2-\overline{\theta}_2\right) -\Sigma_\pi^{(3)}(t)+\Sigma_0^{(3)}(t) \end{equation}  with local boundary fields $\Sigma_{0,\pi}^{(3)}(t)$.  The calculational strategy to determine these fields and the corresponding  differential equations for $F,G$ and the boundary potential $B$ is explained in detail in \cite{mattik2}. We omit the details  here and present only the explicit form of the boundary current $\Sigma_\pi(t)$ along the Neumann directions.  It is given by  
\begin{eqnarray}\nonumber \Sigma^{(3)}_\pi(t) &=& 4m^2\partial_{\io}\partial_iB z^i\zo^{\io}+2m^2\partial_{\io}\partial_{\jo}B \zo^{\io}\zo^{\jo}+2m^2\partial_i\partial_jB z^iz^j \\ \nonumber &\;& +8 \partial_i\partial_{\jo}B \pt z^i\pt\zo^{\jo}+4\partial_{\io}\partial_{\jo}B \pt\zo^{\io}\pt\zo^{\jo} +4\partial_i\partial_jB \pt z^i\pt z^j \\ \nonumber &\;& + 6m^2i\overline{\theta}^{\io}_+\theta_-^i-6im^2\overline{\theta}^{\io}_-\theta_+^i  +8i \pt\overline{\theta}^{\io}_-\pt\theta_+^i -8i\pt\overline{\theta}^{\io}_+\pt\theta^i_+ \\ &\;&  + 4mie^{i\beta} \left(\theta_+^i\pt\theta_+^i-\theta_-^i\pt\theta_-^i\right)  -4mie^{-i\beta} \left(\pt\overline{\theta}_+^{\io}\overline{\theta}^{\io}_+-\pt\overline{\theta}^{\io}_-\overline{\theta}^{\io}_-\right). \end{eqnarray}
The conservation of a higher spin current like (\ref{spin3b})  leads to  strong evidence for the integrability of the underlying field theory, but does clearly not constitute a proof.  As mentioned in section \ref{braneintegrability} one might for the present model furthermore test the mode expansions and commutation relations  of section \ref{openstring} against the requirements derived in \cite{ghoshal} for an integrable boundary theory. These  are in particular the boundary Yang-Baxter equation, the unitarity requirement and  the crossing symmetry which relates the open string mode identifications to the corresponding closed string gluing conditions by an analytic continuation in the so called rapidity variable.  We will  not spell out the details here, but mention that the  modings  derived in section \ref{openstring} fulfil all the requirements presented in \cite{ghoshal}. One  might  compare this  also with the treatment of  the massive Ising model in \cite{ghoshal} and \cite{chatterjee}.\\ Finally,  we want to comment on the number of boundary parameters in the Lagrangian (\ref{boundlagra})  in case of integrable and supersymmetry preserving boundary conditions along a single Neumann direction. From (\ref{zwischen6})  and (\ref{bintegrable1}) in section \ref{openstring} we have  three real parameters as obtained to first order in the bulk coupling constant  for the  ${\cal N}=2$ sine-Gordon model in \cite{nepo2}.  For the sine-Gordon model a calculation taking into account all order contributions reduces this number to a single boundary parameter  as shown in \cite{mattik2}. In the case of present interest, however, contributions leading to these additional constraints  vanish in the $\omega\rightarrow 0$ limit, compare especially with  the quadratic form  of (\ref{bintegrable1}) in comparison with the trigonometric boundary potential  in \cite{nepo2, mattik2} and the discussion in section 4.3 of \cite{mattik2}.

\section{Quantisation}
\label{appquantisation} In this appendix  we  supply some details of the quantisation process  omitted beforehand  in section \ref{openstring}. The required relations in the quantum theory are given by \begin{eqnarray} \label{cano1} \left[x^r(\tau, \sigma), p^s(\tau,\overline{\sigma})\right]&=& 4\pi i \delta^{rs} \delta(\sigma-\overline{\sigma})\\   \label{cano2} \left[x^r(\tau, \sigma), x^s(\tau,\overline{\sigma})\right]&=& 0\\ \left[p^r(\tau, \sigma), p^s(\tau,\overline{\sigma})\right] &=& 0 \label{cano3}\end{eqnarray} for the bosons and \begin{eqnarray} \label{cano4}\left\{ \psi_+^a(\tau,\sigma),\psi^b_+(\tau,\overline{\sigma})\right\} &=& 2\pi\delta^{ab} \delta(\sigma-\overline{\sigma}) \\  \label{cano5}\left\{ \psi_-^a(\tau,\sigma),\psi^b_-(\tau,\overline{\sigma})\right\} &=& 2\pi\delta^{ab} \delta(\sigma-\overline{\sigma})\\ \left\{ \psi_+^a(\tau,\sigma),\psi^b_-(\tau,\overline{\sigma})\right\} &=& 0 \label{cano6} \end{eqnarray} for the fermions,  always understood to be evaluated for $0<\sigma,\overline{\sigma}<\pi$.  Choosing appropriate normalisations of the nonzero modes in the field expansions,   the corresponding commutation relations take on the canonical form presented in section \ref{dirquantisation} and \ref{neuquantisation}. The relations for the zero modes are deduced from that  by using the contour integral method sketched in section \ref{boundaryfermions}.\\ For the fermions along  Dirichlet directions with $a,b\in{\cal D}_-$ we  have  for example \begin{eqnarray} \nonumber \left\{ \psi_+^a(\tau,\sigma),\psi^b_+(\tau,\overline{\sigma})\right\} &=& \left\{\psi^a,\psi^b\right\} e^{-\mt(\sigma+\overline{\sigma})}+\sum_{r\neq 0} e^{ir(\sigma-\overline{\sigma})} +2i\sum_{r\neq 0} c_r^2 \frac{r+i\mt}{\omega_r}\frac{\omega_r-r}{\mt} e^{ir(\sigma+\overline{\sigma})} \\   &=& \left\{\psi^a,\psi^b\right\} e^{-\mt(\sigma+\overline{\sigma})}+\sum_{r\in\mathbb{Z}} e^{ir(\sigma-\overline{\sigma})}+\sum_{r\in\mathbb{Z}} \frac{i\mt}{r-i\mt} e^{ir(\sigma+\overline{\sigma})}\end{eqnarray} with \begin{eqnarray} \sum_{r\in\mathbb{Z}} \frac{i\mt}{r-i\mt} e^{ir(\sigma+\overline{\sigma})}=- \oint_{\cal C} dz \frac{ e^{iz(\sigma+\overline{\sigma})}}{1-e^{2\pi i z}} \frac{i\mt}{z-i\mt}=\frac{-2\pi \mt e^{-\mt(\sigma+\overline{\sigma})}}{1-e^{-2\pi \mt }}\end{eqnarray}  and \begin{eqnarray} \sum_{r\in\mathbb{Z}} e^{ir(\sigma-\overline{\sigma})}=2\pi \delta(\sigma-\overline{\sigma}), \;\;\; 0<\sigma,\overline{\sigma}<\pi. \end{eqnarray}    By using the zero mode anticommutators  (\ref{dirmodesfer1}) one obtains from that the required result (\ref{cano4}).\\[1ex] For the bosons along a Neumann direction we have analogously 
 \begin{eqnarray} \left[x^I(\tau, \sigma), p^J(\tau,\overline{\sigma})\right] &=& \nonumber
2\left(-\left[P^{\Ih}, Q^J\right]+\left[Q^{\Ih}, P^J\right]\right)
\sqrt{(\alphat^{\Ih})^2-\mt^2}e^{\alphat^{\Ih}(\sigma+\overline{\sigma})} 
\\&\;&+2i\delta^{IJ} \sum_{r\in  \mathbb{Z}\backslash\{ 0\}} e^{ir (\sigma-\overline{\sigma})}+2i\delta^{\Ih J} \sum_{r\in  \mathbb{Z}\backslash\{ 0\}} e^{ir (\sigma+\overline{\sigma})}\frac{r-i\alphat^{\Ih}}{r+i\alphat^{\Ih}}\end{eqnarray}
with  \begin{eqnarray}  \sum_{r\in  \mathbb{Z}} e^{ir (\sigma+\overline{\sigma})}\frac{r-ib}{r+ib} =- \oint_{\cal C}dz \;\frac{e^{iz(\sigma+\overline{\sigma})}}{1-e^{2\pi i z}} \frac{z-ib}{z+ib}=4\pi b \frac{e^{b(\sigma+\overline{\sigma})}}{1-e^{2\pi b}}.\end{eqnarray}   From (\ref{cano1}) we deduce (\ref{neumodesbos1}).
\\ Finally, for a fermionic field spanning along a Neumann direction with $I,J\in {\cal N}_-$ we obtain the equation   \begin{eqnarray}\nonumber \left\{\psi^I_+(\tau,\sigma),\psi^J_+(\tau,\overline{\sigma})\right\}&=& \left\{\psi^I,\psi^J\right\} e^{-\mt(\sigma+\overline{\sigma})}+2\left\{\chi^{\Ih},\widetilde{\chi}^J\right\}\frac{\sqrt{(\alphat^{\Ih})^2-\mt^2}+\alphat^{\Ih}}{\mt} e^{\alphat^{\Ih}(\sigma+\overline{\sigma})}\\ &\;\;& +\delta^{IJ}\sum_{r\neq 0} e^{ir(\sigma-\overline{\sigma})}-i\mt\delta^{\Ih J}\sum_{r\neq 0} \frac{1}{r-i\mt}\frac{r-i\alphat^{\Ih}}{r+i\alphat^{\Ih}} e^{ir(\sigma+\overline{\sigma})} \end{eqnarray}
with in this case   \begin{eqnarray}\nonumber  -im\sum_{r\in\mathbb{Z}} \frac{1}{r-im}\frac{r-ib}{r+ib} e^{ir(\sigma+\overline{\sigma})}= \frac{2\pi m}{1-e^{-2\pi m}}\frac{m-b}{m+b} e^{-m(\sigma+\overline{\sigma})}+\frac{4\pi m}{1-e^{2\pi b}}\frac{b}{b+m}e^{b(\sigma+\overline{\sigma})}\end{eqnarray} confirming (\ref{neumodesfer1}) and (\ref{neumodesfer2}). \\ All other relations are either implied by the presented results or are established analogously.

\end{appendix}

\end{document}